%% file: LBGIV_refv2.1.tex
\documentclass[useAMS,usenatbib]{mn2e}

\usepackage{color}
\usepackage{multirow}         
\usepackage{graphicx,epsfig}
\usepackage{multicol}
\usepackage{subfigure}
\usepackage{amsmath,amssymb}

\input{format}

\def\be{\begin{equation}}
\def\ee{\end{equation}}
\def\fltrans{\left<T(s)\right>}
\def\fltranr{\left<T(r)\right>}
\def\vdisp{\left<w_z^{2}\right>^{1/2}}
\def\hmodot{~h^{-1}\rm{M_\odot}}
\def\kps{{\rm km~s}^{-1}}

\def\lowmass{$M_\star>10^8\hmodot$}
\def\lintc{0.11}
\def\lrzero{2.41}

\def\lgam{1.52}

\def\lxibar{0.33}
\def\lxibare{0.02}
\def\lbias{1.85}
\def\lbiase{0.12}
\def\lbeta{0.53}
\def\lbetae{0.07}
\def\lkaiser{1.41}

\def\lkairs{1.26}
\def\lkairse{0.03}

\def\himass{$M_\star>10^9\hmodot$}
\def\hintc{0.21}
\def\hrzero{4.16}

\def\hgam{1.56}

\def\hxibar{0.75}
\def\hxibare{0.05}
\def\hbias{2.80}
\def\hbiase{0.18}
\def\hbeta{0.35}
\def\hbetae{0.02}
\def\hkaiser{1.26}

\def\hkairs{1.21}
\def\hkairse{0.06}

\title[VLT LBG Redshift Survey IV]{The VLT LBG Redshift Survey - IV. Gas and galaxies at $z\sim3$ in observations and simulations}
\author[P. Tummuangpak et al.]{P. Tummuangpak$^{1}$, R. M. Bielby$^{1}$\thanks{E-mail:
rmbielby@gmail.com}, T. Shanks$^{1}$, T. Theuns$^{2,3}$, N. H. M. Crighton$^{4}$,
\newauthor  H. Francke$^{5}$, L. Infante$^{5}$ \\
$^{1}$Department of Physics, University of Durham, South Road, Durham DH1 3LE, UK\\
$^{2}$Institute for Computational Cosmology, Department of Physics, University of Durham, South Road, Durham, DH1 3LE, UK\\
$^{3}$Department of Physics, University of Antwerp, Campus Groenenborger, Groenenborgerlaan 171, B-2020 Antwerp, Belgium\\
$^{4}$Max Planck Institute for Astronomy, K\"{o}nigstuhl 17, D-69117 Heidelberg, Germany\\
$^{5}$Departamento de Astronom\'\i a y Astrof\'isica, Pontificia Universidad Catolica de Chile, Casilla 306, Santiago 22, Chile
}

\begin{document}
\date{}
\pagerange{\pageref{firstpage}--\pageref{lastpage}} \pubyear{2012}

\maketitle
\label{firstpage}
\begin{abstract}

We use a combination of observations and simulation to study the
relationship between star-forming galaxies and the intergalactic medium
at $z\approx3$. The observed star-forming galaxy sample is based on
spectroscopic redshift data taken from a combination of the VLT LBG
Redshift Survey (VLRS) data and Keck LRIS observations in fields centred
on bright background QSOs, whilst the simulation data is taken from
GIMIC. In the simulation, we find that the dominant peculiar
velocities are in the form of large-scale coherent motions of gas and
galaxies. Gravitational infall of galaxies towards one another  is also
seen, consistent with expectations from linear theory. At smaller
scales, the rms peculiar velocities in the simulation overpredict the
difference between the simulated real- and $z$-space galaxy correlation
functions. Peculiar velocity pairs with separations smaller than 1
$h^{-1}$Mpc have a smaller dispersion and explain the $z$-space
correlation function better. The Ly$\alpha$ auto- and cross-correlation
functions in the GIMIC simulation appears to show infall smaller than
implied by the expected $\beta_{Ly\alpha}\approx1.3$ (McDonald et al.).
There is a possibility that the reduced infall may be due to the galaxy
wide outflows implemented in the simulation.

The main challenge in comparing these simulated results with  the
observed Keck $+$ VLRS correlation functions  comes from the presence of
velocity errors for the observed LBGs which dominate at $\la 1h^{-1}$Mpc
scales. When these are taken into account, the observed LBG
correlation functions are well matched by the high amplitude of clustering shown by
higher mass ($M_*>10^9M_\odot$) galaxies in the simulation. The
simulated cross-correlation function shows similar neutral gas densities
around galaxies as are  seen in the observations. The simulated and
observed  Ly$\alpha$ $z$-space autocorrelation functions again agree
better with each other  than with the $\beta_{Ly\alpha}\approx1.3$
infall model.  Our overall conclusion is that, at least in the
simulation, gas and galaxy peculiar velocities  are generally towards
the low end of expectation.  Finally, little direct evidence is seen in
either simulation or observations for high transmission near galaxies
due to feedback, in agreement with previous results.

\end{abstract}

\begin{keywords}
	galaxies: high$-$redshift $,$ intergalactic medium
\end{keywords}

\section{Introduction}
\label{sec:intro}

The effect of feedback via supernovae and AGN driven winds is thought to be a key factor in the process of galaxy formation and evolution. Cosmological models of galaxy formation require efficient injection of feedback from supernovae (SNe) and active galactic nuclei (AGN) to regulate the star formation activity and thus replicate the observed galaxy stellar mass function \citep[e.g.][]{white1978,White1991}. Similarly, cosmological simulations, for example \cite{Springel-Hernquist2003}, \citet{Schaye2010} and \citet{Scannapieco2012}, have shown that supernova feedback is fundamental to recreating the cosmic star-formation history. It is also evident that simulations lacking some sort of feedback struggle to reproduce realistic disk galaxies \citep[e.g.][]{Weil1998,Schaye2010,2012MNRAS.427..379M,Scannapieco2012} and that powerful galactic winds are required to produce the observed metal enrichment of the IGM \citep[e.g.][]{Cen1999,Theuns2002Winds,Aguirre2005,Oppenheimer2006}.

In terms of observing the effects of feedback at high redshift, \citet[][A03 hereafter]{Adelberger2003} presented the cross-correlation between $z\sim3$ galaxies and the IGM (as traced by quasar sightlines) and claimed an observed lack of absorbing gas within $\sim0.5~h^{-1}$Mpc. They interpreted this as evidence of strong galactic winds removing H{\sc{i}} gas from the vicinity of these star-forming galaxies. The work was based on the Keck HiRES ($R\sim40,000$) spectra of 8 background quasars at $z\approx3$ combined with 431 Lyman Break Galaxies (LBGs) from the survey of \citet{Steidel2003}. Following the results of A03, \citet[][A05 hereafter]{Adelberger2005} updated the result with greater numbers of galaxies, this time centred at $z\sim2$. Based on this new sample, A05 found an increase in Ly$\alpha$ absorption down to scales of r $\sim$ 0.5~$h^{-1}$~Mpc of LBG positions, with no evidence for H{\sc i} gas having been removed from the vicinity of these galaxies. Indeed, \citet{Crighton2011} surmised that the cross-correlation at such small scales would likely be affected by uncertainties in the galaxy redshifts in the A03 data. It is therefore still unclear to what extent galactic winds have an effect on this probe of the galaxy surroundings.

In addition to the above evidence for gas outflows, gas infall down to galaxy scales is also predicted in models of galaxy formation \citep[e.g.][]{white1978,2005MNRAS.363....2K,2006MNRAS.368....2D,2009MNRAS.395..160K,Dekel2009,2012MNRAS.421.2809V}. Gas inflow is expected to be coherent  down to the virial radius of a massive galaxy ($\approx140$ kpc), below which scale the situation is more complicated due to shocks and the gas pressure becoming more important. Gas flow infall into galaxies along filaments is also expected in secular models of galaxy formation where the gas accretion rate may not be simply dictated by merging rates in a hierarchical model (\citealt{Dekel2009}). \cite{Rakic2012} presented a study of the galaxy-Ly$\alpha$ cross-correlation at  $z \approx 2.4$ using 15 fields of the Keck Baryonic Structure Survey (KBSS). They saw fingers-of-god on sub-500 kpc scales and evidence for infall on $\sim8$ Mpc scales.

In order to constrain models of galaxy formation, it is imperative to provide extensive observations of the IGM via hydrogen and metal absorption lines and thus identify and probe the infall and outflow processes. As such, we are undertaking a large galaxy survey centred on distant bright quasars in the form of the VLT LBG Redshift Survey (VLRS). \citet{Bielby2011} presented the first stage of the galaxy survey, comprising $\approx$ 1,000 $z\sim3$ galaxies within $\sim30'$ of $z>3$ quasars. Using this sample, \cite{Crighton2011} performed a cross-correlation analysis between the galaxy positions and the Ly$\alpha$ forest of the available quasar spectra in the fields, finding increased absorption within $\sim5~h^{-1}$Mpc of galaxy positions. This result was consistent with the results of A03 and A05, but lacked the galaxy numbers to probe the $\sim0.5~h^{-1}$Mpc scales at which A03 claimed to see the effects of galaxy winds. Since then, the VLRS has been extended to incorporate $\sim2000$ LBGs within 9 separate fields containing bright $z>3$ quasars \citep{2013MNRAS.430..425B}, comparable in number to the only other equivalent surveys at this redshift \citep[e.g.][]{Rakic2012,Rudie2012}.

A number of authors have provided complimentary analysis of such galaxy-gas correlations at $z\sim3$ using smoothed particle hydrodynamical simulations \citep[e.g.][]{2002ApJ...580..634C,Kollmeier2003,2003MNRAS.343L..41B,Desjacques2004,2006MNRAS.367L..74D,2013MNRAS.433.3103R}. Partly prompted by the first survey of LBGs in bright quasar fields, \citet{2002ApJ...580..634C} and \citet{Kollmeier2003} both investigated the possible explanations of the enhancement in the gas profile around LBGs reported by A03, and the distribution of gas around high redshift galaxies in general, using SPH simulations. \citet{2002ApJ...580..634C} found that the absorption profiles around high redshift galaxies increases monotonically with decreasing distance from the galaxies in their simulations. Similarly, \citet{Kollmeier2003} presented consistent results with \citet{2002ApJ...580..634C} showing that, based on their SPH simulations, photoionisation cannot explain the observed reduction in absorption presented by A03.

More recently, \citet{2013MNRAS.433.3103R} used the OverWhelmingly Large Simulation (OWLS) comparing analysis of OWLS to their own observational results \citep{Rakic2012}. As with previous simulation work the authors find a continuous increase in absorption with decreasing distance from a galaxy, consistent with their observations. They go on to analyse the 2-D H{\sc i} Ly$\alpha$ absorption profile and claim a good match between their observations and the simulation, with the gas distribution on scales of $\sim 8$ Mpc being consistent with large-scale gas infall into the potential wells occupied by galaxies.

In this paper, we update the work of \citet{Crighton2011}, adding the galaxy redshifts of \citet{Bielby2011} and also \citet{Steidel2003} in conjunction with the available high-resolution quasar spectra in these survey fields. This work thus combines the higher galaxy sampling rate of the \citet{Steidel2003} survey with the wide fields of the VLRS and provide a galaxy sample that can probe the full range of scales from a few hundreds of kpc to tens of Mpc. This large range of scales is imperative for distinguishing between models of gas inflow and outflow in 2-D galaxy-Ly$\alpha$ cross-correlation analysis. In addition to extending on the previous observational results, we also incorporate a hydrodynamical simulation, the Galaxies-Intergalactic Medium Interaction Calculation \citep[GIMIC, ][]{Crain2009}, into our analysis in order to interpret the observations.

This paper is organised as follows. Observational data from the VLT LBG Redshift Survey and Keck LBG observations of \citet{Steidel2003} are described in section 2. Section 3 describes the GIMIC simulations. The simulated galaxy clustering results and their interpretation are  shown in section 4, while the galaxy-IGM cross-correlation is presented in section 5. Section 6 presents an analysis of the Ly$\alpha$ auto-correlation in both the observations and the simulation. Our discussion and conclusions are presented in section 7 and 8 respectively.
	
Throughout this work, we adopt a cosmology consistent with the GIMIC simulation (and hence the Millennium  simulation, \citealt{2005Natur.435..629S}). This corresponds to $\{\Omega_m,\Omega_\Lambda,\Omega_b,n_{\rm s},\sigma_8,H_0,h\}=\{0.25,0.75,0.045,1,0.9,100,0.73\}$. As we are working in both real and redshift space in this paper, it is prudent to note the conventions on coordinates that we use here. For real-space separations between two points, we use $r$, whilst in redshift space, we use $s$. Where a plot shows results in both real and redshift space (i.e. where we show simulation results), we denote the distance axis with $r$. For the observed data, all distances are of course measured in redshift space and so separations are denoted by $s$ in any plots primarily showing observational data. We denote the transverse and line of sight coordinates with $\sigma$ and $\pi$ respectively, regardless of whether these are in real or redshift space. All coordinates are given in comoving coordinates in this paper unless stated otherwise.

\section{Observations}

In this work, we use a combination of spectroscopically identified $z\sim3$ star-forming galaxies and high-resolution spectral observations of the Ly$\alpha$ forest of $z \gtrsim 3$ quasars. The galaxy data are a combination of the VLRS data presented by \citet{Bielby2011} and \citet{2013MNRAS.430..425B}, and the publicly available Keck LBG data presented by \citet{Steidel2003}. These two datasets are based on different observing strategies, whereby the VLRS offers coverage across large fields of view, whilst the Keck sample covers relatively small separations ($\lesssim8-10$ Mpc) with higher sampling rates of the galaxy population. The quasar spectra with which we trace the distribution of H{\sc i} within the fields have all been obtained from archival VLT UVES and Keck HiRES observations. In this section, we give details of all the data and the reduction processes used for the quasar spectra.

\subsection{LBG Observations}

The VLRS currently provides $\sim2,000$ spectroscopic galaxy redshifts within 9 fields centred on $z \gtrsim 3$ quasars \citep{Bielby2011,2013MNRAS.430..425B}. The redshifts were obtained using the VLT VIMOS instrument \citep{2003SPIE.4841.1670L} with the LR\_Blue grism, giving a resolution of $R\sim180$ and velocity accuracies of $\sigma_v\approx350~\kps$. In total, the survey covers an area of $\sim2.6$ deg$^{2}$ and provides galaxy data in the foreground of the following 9 high redshift quasars: Q0042-2627 ($z=3.29$), J0124+0044 ($z=3.84$), Q0301-0035 ($z=3.23$), HE0940-1050 ($z=3.05$), J1201+0116 ($z=3.23$), PKS2126-158 ($z=3.28$), Q2231+0015 ($z=3.02$), Q2348-011 ($z=3.02$) and Q2359+0653 ($z=3.23$). The spectroscopic galaxy sample is predominantly limited to $R<25$ (Vega) although a number of fainter galaxies ($R<25.5$ Vega) are present in the sample where slit allocation during the VIMOS observations could be optimised by their inclusion.

The LBG redshifts were identified using Ly$\alpha$ emission lines and interstellar medium (ISM) absorption lines where visible. For both the Ly$\alpha$ and ISM features, it is necessary to correct the measured redshift for intrinsic velocity effects, due to these features being affected by outflowing gas \citep[e.g. A03,][]{Steidel2010}. As such the VLRS galaxy redshifts have been corrected according to the prescription given by \citet{Steidel2010}.

The Keck survey provides a sample of $\sim$ 940 LBGs observed using the Keck LRIS instrument \citep{lris}. The quasars from six (Q0201+1120, Q0256-0000, Q0302-0019, B0933+2854, Q2233+1341 and Q1422+2309) out of the 17 Keck fields are available to us through the public archive and taking only those galaxies in fields around these 6 Keck quasars, the numbers of LBGs are reduced to 308. The Keck LBGs are limited to ${\cal R}=25.5$ (AB).

\subsection{Quasar data}

We have analysed publicly available archival spectroscopy for 16 quasars in
the redshift range $2.9\lesssim z\lesssim3.6$, with an additional quasar spectra provided by our own X-Shooter observations to make a total of 17 quasar sightlines. The publicly available data are all high resolution ($R\gtrsim30,000$), high signal-to-noise ($S/N\gtrsim20$) spectra observed using either the UVES instrument \citep{uves} on the VLT or the HiRES instrument \citep{hires} on the Keck telescope. Full details of the reduction of UVES and HiRES quasar spectra for 11 of the quasars used
here are provided by \citet{Crighton2011}. The remaining 6 spectra were
all observed with the Keck HiRES instrument and were reduced following
an identical method to that used for the two Keck quasars of
\citet{Crighton2011}, using the \textsc{makee} package\footnote{www2.keck.hawaii.edu/inst/common/makeewww}. Briefly, this encompassed basic flat-fielding and bias subtraction,
followed by the use of \textsc{spim2} to splice the echelle orders and
combine individual observations. This involved producing template
spectra constructed by combining the individual observations, masking
bad regions of the CCDs and rescaling. A template was applied to rescale
the original observations. We divide out the continuum for each
individual observation, then multiply this normalised flux by a
continuum fit to the template. After scaling each order of each
observation individually, we combined them to get the final spectrum.

In addition to the publicly available quasar spectra, we also include a spectrum from our own observations using the X-Shooter instrument \citep{xshooter} on the VLT for the quasar Q2359+0653. This data was reduced using the X-Shooter pipeline package - see Bielby et al. (in prep) for details. The full list of quasars used in this study is provided in Tab.~\ref{tab:qsolist}.

	\begin{table*}
	\caption{List of quasars used in this study.}
	\centering
	\begin{tabular}{lcclll}
	\hline
	Quasar & R.A. & Dec. & $z$ & Mag & Instrument \\
    	& \multicolumn{2}{c}{J2000} & & & \\
	\hline
	Q2359+0653      & 00:01:40.6 & +07:09:54 	& 3.23  & $V=18.5$   & X-Shooter\\
	Q0042-2627      & 00:44:33.9 & -26:11:19 	& 3.289 & $B=18.5$   & HIRES \\
	WHO91 0043-265  & 00:45:30.5 & -26:17:09    & 3.44  & $R=18.3$   & HIRES \\
	J0124+0044      & 01:24:03.8 & +00:44:32 	& 3.83  & $g=19.2$   & UVES  \\
	Q0201+1120      & 02:03:46.7 & +11:34:45	& 3.610 & $G=20.1$   & HIRES \\
	Q0256-0000      & 02:59:05.6 & +00:11:22	& 3.364	& $G=18.2$   & HIRES \\
	Q0301-0035      & 03:03:41.0 & -00:23:22 	& 3.230 & $g=17.6$   & HIRES \\
	Q0302-0019      & 03:04:49.9 & -00:08:13 	& 3.281	& $G=17.8$   & HIRES \\
	B0933+2845      & 09:33:37.2 & +28:45:32 	& 3.428 & $G=17.5$   & HIRES \\
	HE0940-1050     & 09:42:53.5 & -11:04:25 	& 3.06  & $B=17.2$   & UVES  \\
	J1201+0116      & 12:01:44.4 & +01:16:11 	& 3.233 & $g=17.7$   & HIRES \\
	Q1422+2309      & 14:24:38.1 & +22:56:01 	& 3.620 & $G=16.5$   & HIRES \\	
	Q2129-1602      & 21:29:04.9 & -16:02:49    & 2.90  & $R=19.2$   & HIRES \\   
	PKS2126-158     & 21:29:12.2 & -15:38:40 	& 3.268 & $V=17.3$   & UVES  \\
	Q2231+0015      & 22:34:08.9 & +00:00:01 	& 3.02 	& $r=17.3$   & UVES  \\	
	Q2233+1341      & 22:36:27.2 & +13:57:13 	& 3.209	& $G=20.0$   & HIRES \\
	Q2348-011       & 23:50:57.9 & -00:52:10 	& 3.023	& $r=18.7$   & UVES  \\
	\hline
	\end{tabular}
	\label{tab:qsolist}
	\end{table*}

\section{GIMIC simulations}

\subsection{Overview}

We simulate both Ly$\alpha$ spectra and galaxies to compare with the observational data using a hydrodynamical cosmological simulation. Our main aims are to study the real and redshift-space auto and cross-correlation functions. We wish to ascertain as to whether we can detect the effects of peculiar velocities in order to understand more about gas outflow and infall around galaxies, for (a) LBG-LBG pairs (b) Ly$\alpha$-Ly$\alpha$ pairs and (c) the LBG-Ly$\alpha$ forest. The results will then be used to interpret the observable 1-D and 2-D correlation functions $\xi(r)$ and $\xi(\sigma,\pi)$ in terms of both simulation and observational results. As outlined earlier, $\sigma$ denotes the distance transverse to the line of sight, $\pi$ denotes the line of sight distance and $r$ is the (real-space) vector combination of the two coordinates, thus $r=\sqrt{\sigma^2+\pi^2}$. When working in redshift space we use $s$ in place of $r$.

We use the GIMIC simulation, which is a cosmological hydrodynamical re-simulation of selected volumes of the Millennium simulation \citep{2005Natur.435..629S}. GIMIC is designed to overcome the issues in simulating large cosmological volumes ($L\gtrsim100h^{-1}$ Mpc) at high resolution ($m_{gas}\lesssim10^7h^{-1}$ M$_\odot$) to $z=0$ by taking a number of smaller regions with `zoomed' initial conditions \citep{Frenk1996,Power2003,Navarro2004}. These individual regions each have approximate radii of $18h^{-1}$Mpc outside of which the remainder of the Millennium simulation volume is modelled with collisionless particles at much lower resolution.

GIMIC was run using the TreePM SPH code GADGET3, which is an update of the GADGET2 code \citep{Springel2005}. The cosmological parameters adopted were: $\Omega_m$ = 0.25, $\Omega_{\lambda}$ = 0.75, $\Omega_b$ = 0.045,~$h_0$ = 100~$h~\kps$Mpc$^{-1}$,~$h$ = 0.73, $\sigma_8=0.9$ and $n_s=1$ (where $n_s$ is the spectral index of the primordial power spectrum).

Radiation cooling and stellar evolution were implemented as described in \citet{Wiersma2009}, whilst star-formation was handled as described by \citet{2008MNRAS.383.1210S} and supernova feedback was implemented following the prescription of \citet{DallaVecchia2008}.

The GIMIC simulations are particularly well suited to the study of $\sim L^\star$ galaxies. As shown in \citet{Crain2009}, the implementation of efficient (but energetically feasible) feedback from SNe largely prevents overcooling on the mass scale of $L^\star$ galaxies, and is key to the reproduction of the observed X-ray scaling relation presented in that study. Indeed, GIMIC accurately reproduces the rotation speeds and star formation efficiencies of $z=0$ disc galaxies for $10^9 \lesssim M<10^{10.5}\rm{M}_\odot$, although galaxies with $M_\star\gtrsim10^{11}\rm{M}_\odot$ do still suffer from some overcooling \citep{2012MNRAS.427..379M}. Moreover, \citet{2011MNRAS.416.2802F} demonstrated that $L^\star$ galaxies in GIMIC exhibit satellite luminosity functions and stellar spheroid surface brightness distributions that are comparable to those of the Milky Way and M31, whilst \citet{2012MNRAS.420.2245M} further demonstrated that this correspondence extends also to their global structure and kinematics.

In terms of reproducing the Ly$\alpha$ forest, \citet{1998MNRAS.301..478T} conducted simulations across a range of resolutions (i.e. gas particle masses) in order to evaluate the effect of resolution on such studies. They found convergence of the mean effective optical depth (at $z=3$) in their SPH simulations at gas particle masses of $\lesssim1.4\times10^8\hmodot$, whilst column density distributions were found to be consistent given gas particle masses of $\lesssim1.8\times10^7\hmodot$. Both of these limits are significantly higher than the GIMIC gas particle mass of $1.45\times10^6\hmodot$ \citep{Crain2009}, indicating that resolution effects are not an issue for our work in terms of the Ly$\alpha$ forest. In terms of the selected DM halos, the dark matter particle masses in GIMIC are $6.6\times10^{6}\hmodot$, which is $\gtrsim2$ orders of magnitude lower than any halo mass we will be considering in this study.

In this work, we focus on the Ly$\alpha$ forest, i.e. $N_{\rm{HI}}\lesssim10^{17}~\rm{cm}^{-2}$. In this regime, the gas is optically thin, such that radiative transfer implementations such as that of \citet{2011ApJ...737L..37A} are not necessary.

An area of interest for this study is the effect of supernovae (SNe) feedback on the local environment of galaxies. GIMIC contains an implementation of SN feedback based on the generation of winds as follows. Firstly, after a delay corresponding to the maximum lifetime of stars that undergo core collapse SNe, newly formed star particles impart a randomly directed 600 km~s$^{-1}$ kick to, on average, $\eta=4$ of its neighbours. Here $\eta$ is the mass loading (defined as $\eta\equiv\overset{.}{m}_{\rm{wind}}/\overset{.}{m}_\star$) and its value for GIMIC was chosen to match the global star formation rate density to observational data. The $600~\rm{km~s}^{-1}$ initial kick is not equivalent to measured outflow velocities given that it is a `launch' velocity and is not necessarily what observations measure. In addition, the particles that receive this wind kick are never decoupled from the hydrodynamical calculations, as is done in e.g. \citealt{Springel-Hernquist2003,Oppenheimer2008,2009MNRAS.395..160K,2013MNRAS.436.2929H}, and so are subject to significant deceleration as they travel into either the host galaxy disk or halo. We also note that the above simulations specifically direct the applied wind kicks perpendicular to the plane of the host galaxy, as opposed to the randomly orientated wind kicks used in GIMIC. Ultimately, the lack of decoupling and the isotropic nature of the wind kicks in GIMIC means that the values of the launch velocity and mass loading used in GIMIC are necessarily higher than in the above studies. We note that the wind launch velocity used in GIMIC is consistent with the higher end of the Ly$\alpha$ wind velocities reported by \citet{2001ApJ...554..981P} and \citet{Shapley2003}, lending the value some legitimacy.

In the work presented here, we use the `$0\sigma$' GIMIC region, which is identified as having a mean density at $z=1.5$ equal to the mean density of the Universe at that epoch. In addition, we use only one snapshot of this region, chosen to be at a redshift of $z=3.1$ in order to provide a suitable comparison to our $z\sim3$ observed population of star-forming galaxies. All the analysis is limited to a sphere of radius $16~h^{-1}$Mpc in order to negate the effects of particles being `moved' out of the analysis region when moved to redshift-space. Given a limiting radius of $16~h^{-1}$Mpc, the same number of \lowmass\ galaxies are present in the region regardless of whether redshift-space distortions (RSD) are applied or not.

\subsection{Simulated galaxy population}

\subsubsection{Identifying the galaxy population}

The galaxy population is identified in the simulation based on first identifying the dark matter halos using a Friends of Friends \citep[FoF,][]{Davis1985} algorithm. A group finding algorithm then locates the nearest dark matter halo for each baryonic (gas or star) particle and identifies the particle with this halo. The {\sc subfind} algorithm \citep{Springel2001,Dolag2009} is then used to identify self-bound sub-structures within the halos, to which star particles are associated and defined as galaxies. 

\begin{figure}
	\centering
	\includegraphics[width=80.mm]{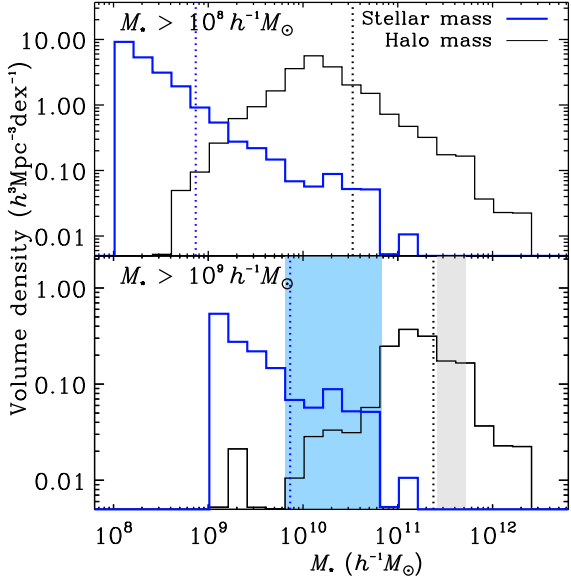}
	\caption{Distribution of total halo and galaxy stellar masses for the two GIMIC galaxy selections, $M_\star>10^{8}\hmodot$ (top) and $M_\star>10^{9}\hmodot$ (bottom). The blue histograms in each panel show the numbers of galaxies as a function of stellar mass, whilst the black histograms show the numbers of galaxies as a function of total halo mass. The dotted vertical lines show the mean halo mass, $M_{\rm halo}=10^{10.5}\hmodot$ and $M_{\rm halo}=10^{11.4}\hmodot$ for the low and high mass cuts respectively. The shaded light blue region in the lower panel shows the observed $1\sigma$ range in stellar masses of $z\approx3$ LBGs from \citet{Shapley2005}.}
	\label{fig:gimic-masses}
	\end{figure}

We use cuts in stellar mass to define our simulated galaxy samples. In the first instance we take galaxies with stellar masses of \lowmass. This is intended as a large sample, which is not representative of the $z\sim3$ population sampled by present observations, but acts as a comparison data-set for a second more representative sample. Taking our limiting radius within the GIMIC volume of 16~$h^{-1}$ Mpc radius, this low-mass cut gives a sample of 4,070 galaxies from the snapshot at $z = 3.06$ in the $0\sigma$ density region. The distribution of galaxy stellar mass (blue histogram) and host halo mass (black histogram) for this sample is shown for reference in the top panel of Fig.~\ref{fig:gimic-masses}. The mean galaxy stellar mass is $M_\star=10^{8.9}\hmodot$ (blue vertical dashed line), whilst the mean host halo mass is $M_{\rm halo}=10^{10.5}h^{-1}\hmodot$ (black vertical dashed line).

With our second simulated galaxy sample, we aim to mimic more closely the observed LBG samples and specifically to reproduce the observed clustering. The \himass\ cut used above provides a simulated galaxy sample with a clustering signal well matched to the observed clustering of LBGs (see section~\ref{sec:autocorr}). \citet{2013MNRAS.430..425B} present measurements of the clustering of the VLRS spectroscopic $z\sim3$ galaxy sample, estimating a clustering length of $r_0=3.83\pm0.24\hmpc$ and typical halo masses of $10^{11.57\pm0.15}\hmodot$. Similarly, \citet{2005ApJ...619..697A} measure $r_0=4.0\pm0.6\hmpc$ and halo masses of $10^{11.5\pm0.3}\hmodot$ for a comparable sample of $z\sim3$ LBGs. We thus vary the stellar-mass constraints on the galaxy selection to match these clustering/mean halo mass  results (where the total masses for the GIMIC galaxies are available from the {\sc subfind} algorithm). We show the clustering results in section~\ref{sec:autocorr}, whilst the resulting stellar and halo mass distributions are shown in the lower panel of Fig.~\ref{fig:gimic-masses}. We find that a stellar mass cut of \himass\ reproduces the observed clustering well and gives a mean halo mass for the simulated galaxies of $M_{\rm{halo}}=10^{11.4}\hmodot$, marginally lower than the observed samples, but consistent at the $\sim1\sigma$ level.

The mean of the galaxy stellar masses is $M_\star=10^{9.9}\hmodot$ (blue vertical dotted line). This \himass\ cut gives a sample of 287 simulated galaxies within $16~h^{-1}$~Mpc of the centre of the GIMIC volume, equating to a space density of $\rho_{g}\sim5\times10^{-3}~h^3$Mpc$^{-3}$ (for comparison, \citealt{2005ApJ...619..697A} measure a space density of $\rho_{g}=4\pm2\times10^{-3}~h^3$Mpc$^{-3}$ for the Keck LBG sample). 

The cyan shaded region in the lower panel of Fig.~\ref{fig:gimic-masses} shows the standard deviation range around the mean galaxy stellar mass derived from the observations of \citet{Shapley2005} (i.e. $M_{\star}=10^{10.32}\hmodot$, with a standard deviation of $\sigma_{{\rm log}(M\star)}=0.51$ dex). The galaxy stellar mass of the GIMIC selection overlaps the range of the observed galaxies, but extends further to lower stellar masses (i.e. $M_\star<10^{9.5}\hmodot$). We note that the \citet{Shapley2005} result is based on $K_s$ observations and that 23\% of their UV selected sample is not included in the stellar mass distribution due to not being detected in the $K_s$ observations. This bias against LBGs fainter in the $K_s$ band means that the \citet{Shapley2005} stellar mass distribution lacks some of the lower-mass population, but is unlikely to explain the entire discrepancy between the GIMIC stellar masses and the observed mean galaxy stellar mass. Further to this \citet{Crain2009} calculate the galaxy stellar mass functions from the GIMIC simulation suite and compare to observations at $z=2$, showing the simulated galaxy mass functions to have a significantly steeper slope at $M_\star\lesssim9-10$. They surmise that this reflects a reduction in the efficiency of SNe feedback in the simulation for low mass galaxies.

We also note that the simulation was performed with a relatively high value for $\sigma_8$ (a value of $\sigma_8=0.9$ which originated from a combined analysis of the Two-degree-Field Redshift Survey (2dFGRS) and the Three-Year Wilkinson Microwave Anisotropy Probe (WMAP3) data, \citealt{2005Natur.435..629S}) when compared to the present observed constraints ($\sigma_8=0.83\pm0.01$, \citealt{2013arXiv1303.5076P}) and so for a given mean halo mass (and galaxy stellar mass), we would expect a higher clustering amplitude from the simulation when compared to the observations. This is indeed seen, as although the mean halo mass and mean galaxy stellar mass are lower in the simulated sample than the observed samples, the clustering amplitude ($r_0=4.16\hmpc$) is marginally higher than the observed $r_0$ values from both \citet{2013MNRAS.430..425B} and A05.

All combined, the GIMIC \himass\ simulated galaxies provide a population that is consistent with the observed LBG population in number density and clustering, although the $M_\star$ profiles extend to somewhat lower stellar masses than observed (at least in $K$-band detected samples).

\subsubsection{Velocity field of the simulated galaxies}

\begin{figure}
	\centering
	\includegraphics[width=90.mm]{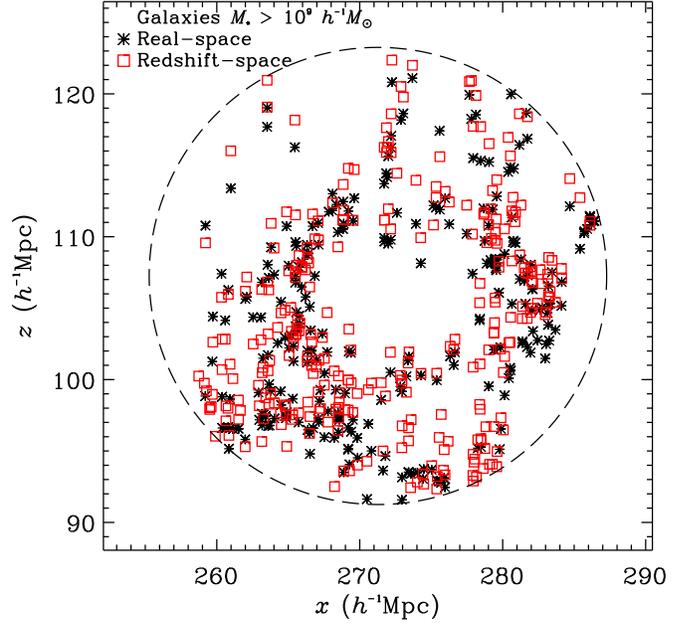}
	\caption{The distribution of \himass\ simulated galaxies in the $x$-$z$ plane in real-space (black asterisks) and redshift-space (open red squares), where the $z$-direction is the redshift/line of sight dimension in this work. The dashed circle shows the volume limit that we place on the simulation data, given by a radius of $16\hmpc$ from the volume centre.}
	\label{fig:galRZ}
	\end{figure}

The  distribution of \himass\ galaxies in real- (black asterisks) and redshift-space (red squares) is shown in Fig.~\ref{fig:galRZ}. Throughout this work, we use the $x$ and $y$ coordinates within the simulation as the transverse to the line of sight coordinates and $z$ as the line of sight coordinate, either in real or redshift-space. Fig.~\ref{fig:galRZ} illustrates the measured positional shifts in the $z$-direction given by the peculiar velocities of the galaxies within the simulation. It is evident from this plot that there is an overall large scale `bulk' motion directed in the positive redshift direction due to the motion of the zoomed region with respect to the full $500~h^{-1}\rm{Mpc}$ Millennium volume. Measuring the distribution of the galaxy velocities, we find an average velocity $\left<v\right> = 93~\kps$ with a standard deviation of 128~km~s$^{-1}$ for the \himass\ galaxy sample and $\left<v\right> = +94~\kps$ with a standard deviation of 125~km~s$^{-1}$ for the \lowmass\ galaxy sample.

	\begin{figure}
	\centering
	\includegraphics[width=80.mm]{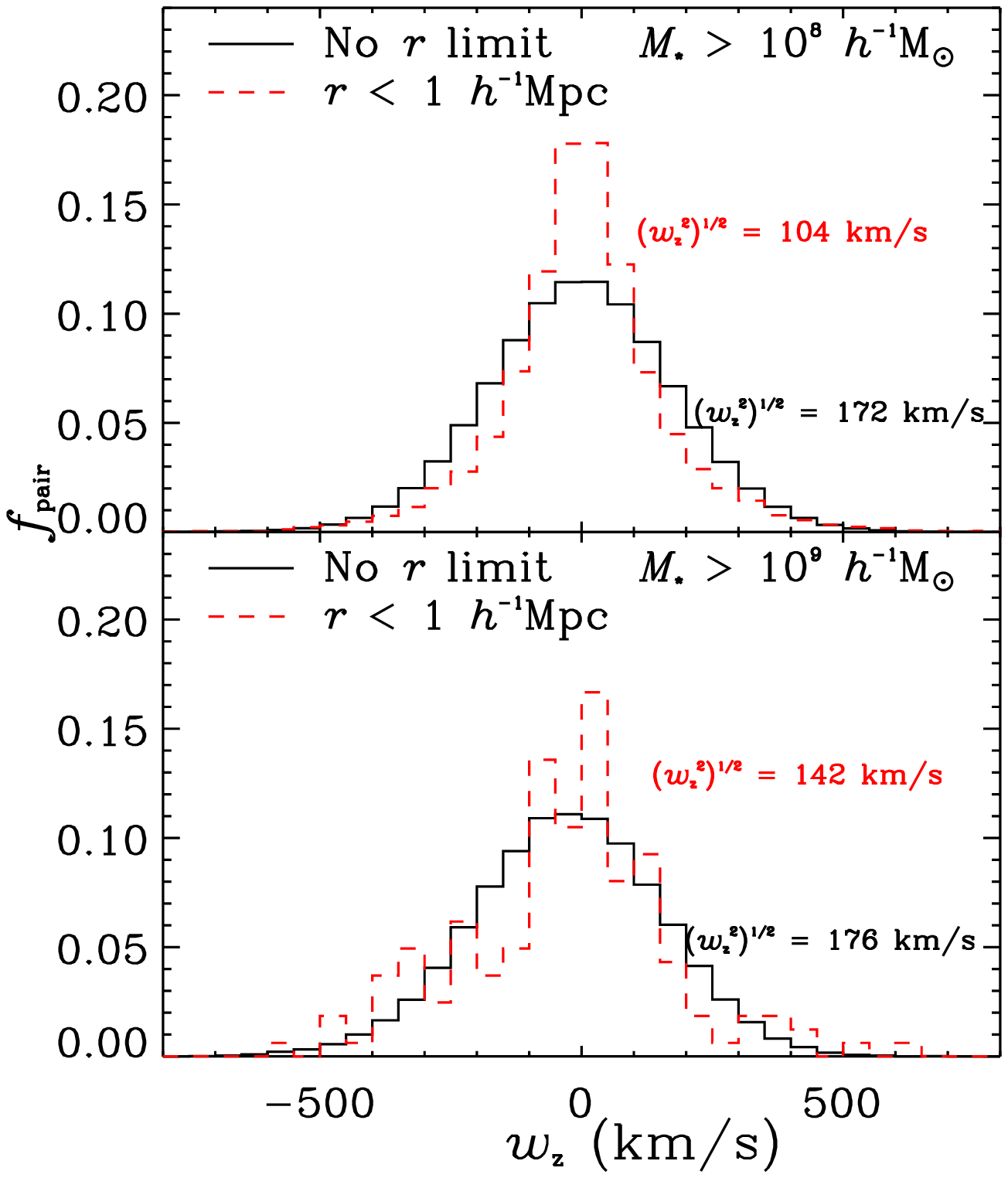}
	\caption{The distribution of pairwise velocities ($w_{\rm z}$, solid histograms) for the GIMIC galaxy samples. The top panel shows the distribution for the \lowmass\ galaxy cut and the lower panel that for the \himass\ cut. Given the effect of pairwise velocities will be dominant at small scales (i.e. $\lesssim1\hmpc$), we also show the distributions in each case for only those pairs separated by $r<1\hmpc$ (dashed red histograms in both panels). The resulting RMS pairwise velocities are indicated in each case and the separation limit gives smaller values of the RMS pairwise velocity in both cases.}
	\label{fig:gal-vz}
	\end{figure}

We show the pairwise velocity, $\vdisp$, distributions (solid black histograms) of galaxies in Fig.~\ref{fig:gal-vz} (where $w_z$ is the line of sight velocity difference between two objects). For the \lowmass\ and \himass\ galaxy samples we find $\vdisp=172~\kps$ and $\vdisp=176~\kps$ respectively. The red dashed histograms show the distribution for only those pairs within $1$~$h^{-1}$Mpc of each other, thus isolating the intra-halo velocity dispersion and excluding the effect of the halo-halo velocity dispersion. This is important when considering the effect of the velocity dispersion on the galaxy-galaxy clustering measurement. The standard deviations of the pairwise velocities for pairs within $1$ $h^{-1}$Mpc are 104~km~s$^{-1}$ and 142~km~s$^{-1}$ for \lowmass\ and \himass\ galaxies respectively. None of these standard deviations include redshift uncertainties due to measurement errors that affect the observed galaxy redshifts.

\subsection{Simulating the Ly$\alpha$ forest spectra}

	\begin{figure}
	\centering
	\includegraphics[width=90.mm]{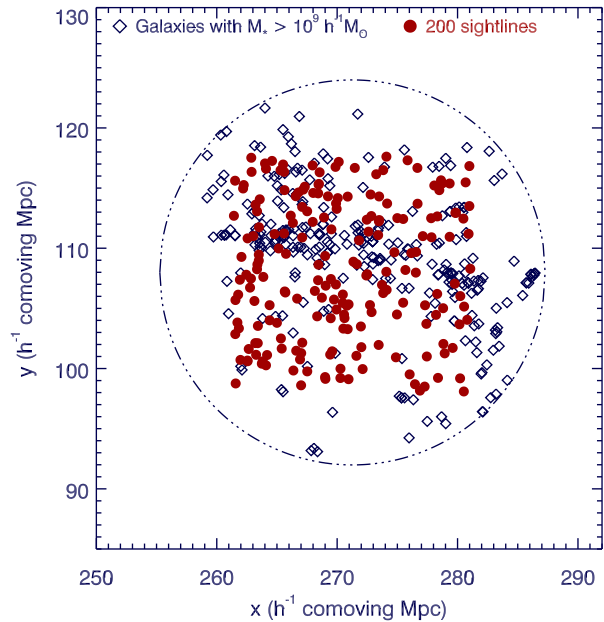} 
	\caption{The position of \himass\ galaxies (diamonds) and 200 Ly$\alpha$ sightlines (circles) projected onto the $x-y$ plane (i.e. equivalent to an on-sky projection).}
	\label{fig:gal_lya_xy}
	\end{figure}

\begin{figure*}
	\centering
	\includegraphics[width=160.mm]{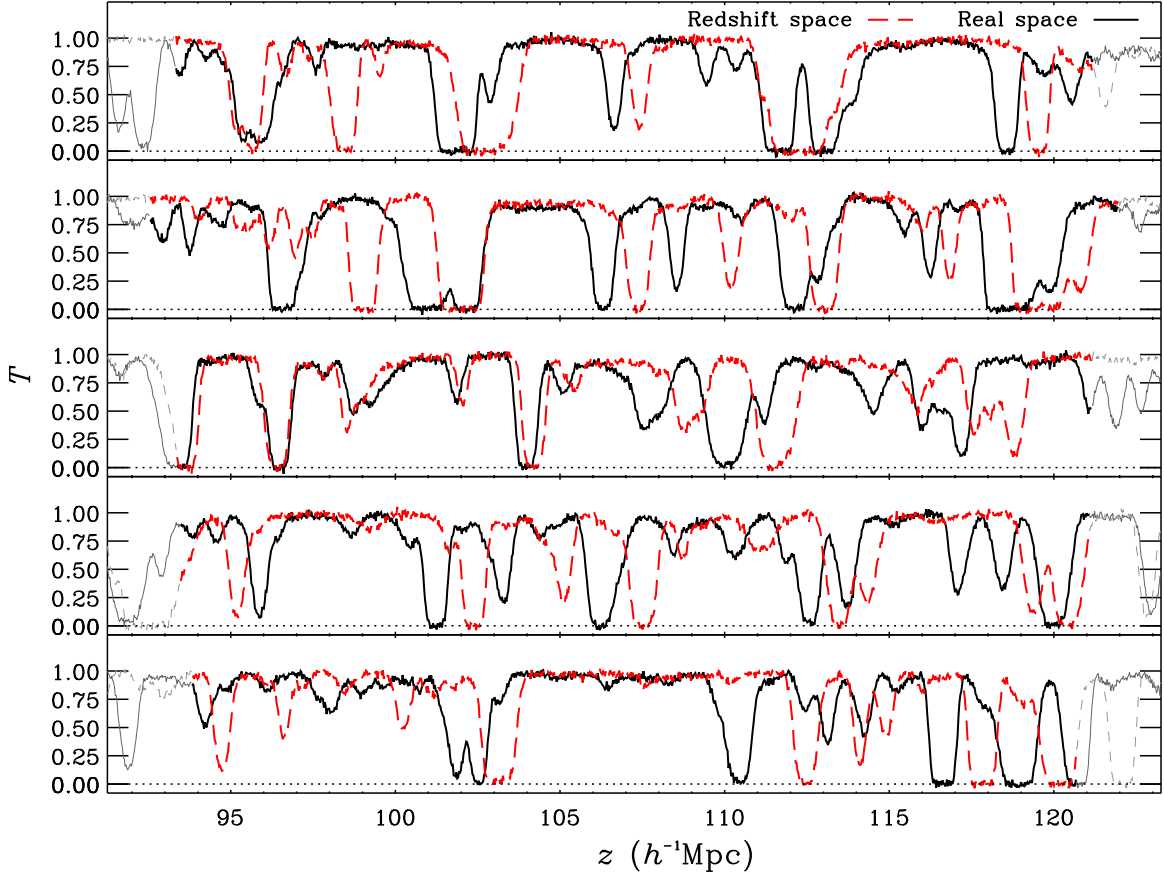} 
	\caption{Examples of absorption spectra from the simulated quasar sightlines. The black solid profiles show $T=e^{-\tau}$ in real-space, whilst the red dashed line shows $T$ in redshift-space. A cut is imposed on the simulated spectra at $r=16\hmpc$ from the centre of the simulation volume. Sections of the spectra that lie outside this sphere are shown in light grey.} 
	\label{fig:spec}
	\end{figure*}

We generate spectra along the $z$-direction through the GIMIC volume. The sightlines were extracted using {\sc specwizard}\footnote{Developed by J. Schaye, C. Booth and T. Theuns, see \citet{1998MNRAS.301..478T} for details}. The transmission is given by, $T = e^{-\tau}$, where $\tau$ is the optical  depth along the line-of-sight. We use a spectral resolution FWHM of 7~km~s$^{-1}$ to convolve each spectrum, a signal-to-noise of 50 per pixel, and pixels of width 2.8~km~s$^{-1}$ which are typical values of our UVES and HIRES quasar spectra. The sightlines were generated parallel to the $z$-axis with random $x$ and $y$ positions. We constructed 200 sightlines with each sightline being constrained not to extend beyond 16~$h^{-1}$ Mpc from the centre of the GIMIC volume in order to avoid any edge effects in terms of the gas extent (see Fig.~\ref{fig:gal_lya_xy}). The average transmission, $\bar T_r$ for real-space is 0.69 while the $\bar T_z$ for redshift-space is 0.72. This difference is likely due to infall of saturated absorption lines towards each other in redshift-space, which results in an overall increase in the measured transmission. This will cause the average transmissivity over the full spectrum to increase in redshift-space as seen. Some hint of this effect can be seen in Fig.~\ref{fig:spec} in which we show a number of examples of the flux from each sightline compared in real (black lines) and redshift (red lines) space. These values for the mean transmission at $z\sim3$ are consistent with the observed values at the $\sim1-2\sigma$ level - for example \citet{McDonald2000} measure a value of $\overset{\_}{T}(z=3)=0.684\pm0.023$, whilst measurements of the effective optical depth by \citet{Faucher-Giguere2008} give $\overset{\_}{T}(z=3)=0.680\pm0.020$.

\begin{figure}
	\centering
	\includegraphics[width=70.mm]{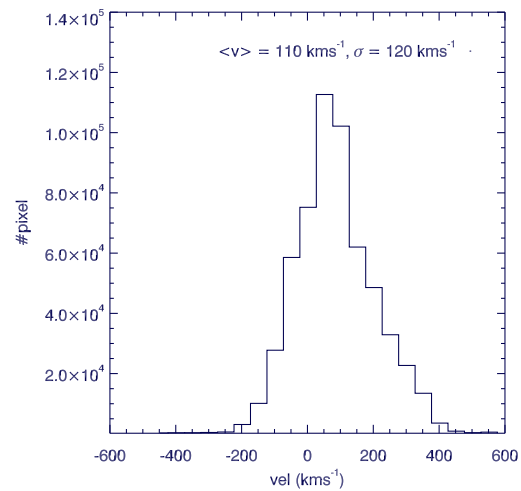}
	\caption{The distribution of LOS optical depth weighted peculiar velocities measured within each pixel in each of the GIMIC simulated spectra. This illustrates the underlying dynamics present in the spectra. The LOS peculiar velocity distribution shows a mean peculiar velocity of $\left<v\right>=110~\kps$ with a standard deviation of 120~km~s$^{-1}$.}
	\label{fig:lya_vel}
	\end{figure}

Using {\sc specwizard}, we calculate the optical depth weighted line-of-sight (LOS) peculiar velocities for each pixel in our 200 spectra. The distribution of the peculiar velocities is given in Fig.~\ref{fig:lya_vel}. As with the galaxy population, the gas traced by the simulated spectra shows the bulk motion in the positive $z$-direction, with a mean peculiar velocity of $\left<v\right> = 110~\kps$ and a standard deviation of 120~km~s$^{-1}$. The standard deviation of the gas peculiar velocity is comparable to that measured for the galaxy samples ($\approx125-130~\kps$).
	
\section{Galaxy Clustering}
\label{sec:autocorr}

\subsection{1-D correlation function}

\subsubsection{Estimator}
\cite{2013MNRAS.430..425B} presented a clustering analysis of the LBG
data used in this study (combining the VLRS and Keck data). In this
section, we compare the observed galaxy clustering presented by
\cite{2013MNRAS.430..425B} to results obtained using the galaxy
population within the GIMIC simulation. In so doing, we may validate how
representative the GIMIC galaxy population is of the observed
$z\approx3$ LBG population in terms of intrinsic clustering properties
and the effects of the galaxy velocity field on the galaxy clustering.

We calculate the real and redshift-space functions, $\xi(r)$ and $\xi(s)$, of the GIMIC $z=3.06$ galaxy samples using the \citet{1983ApJ...267..465D} estimator:

 	\begin{equation}
          \xi(r)  =  \frac{N_R}{N_G}\frac{\left<DD(r)\right>}{\left<DR(r)\right>} - 1,
 	 \end{equation}	

\noindent where $\left<DD(r)\right>$ is the average number of galaxy-galaxy pairs and $\left<DR(r)\right>$ is the number of pairs of galaxy-randoms at the separation, $r$. The factor $\frac{N_R}{N_G}$ is the ratio of the number of random to
data points.
 		
We estimate errors on the auto-correlation results using jack-knife estimates based on splitting the simulation into equal volume octants and excluding each octant in turn to create 8 jack-knife realisations of the data. The correlation functions are then fit using a power-law of the form of:

\begin{equation}
\label{eq-power-law}
\xi(r)=\left(\frac{r}{r_0}\right)^{-\gamma},
\end{equation}
	
\noindent where $\gamma$ is the slope of clustering, $\xi(r)$, and 
${r_0}$ is the clustering length.
	
\subsubsection{Simulated real-space galaxy correlations}

\begin{figure*}
\centering
\includegraphics[width=88.mm]{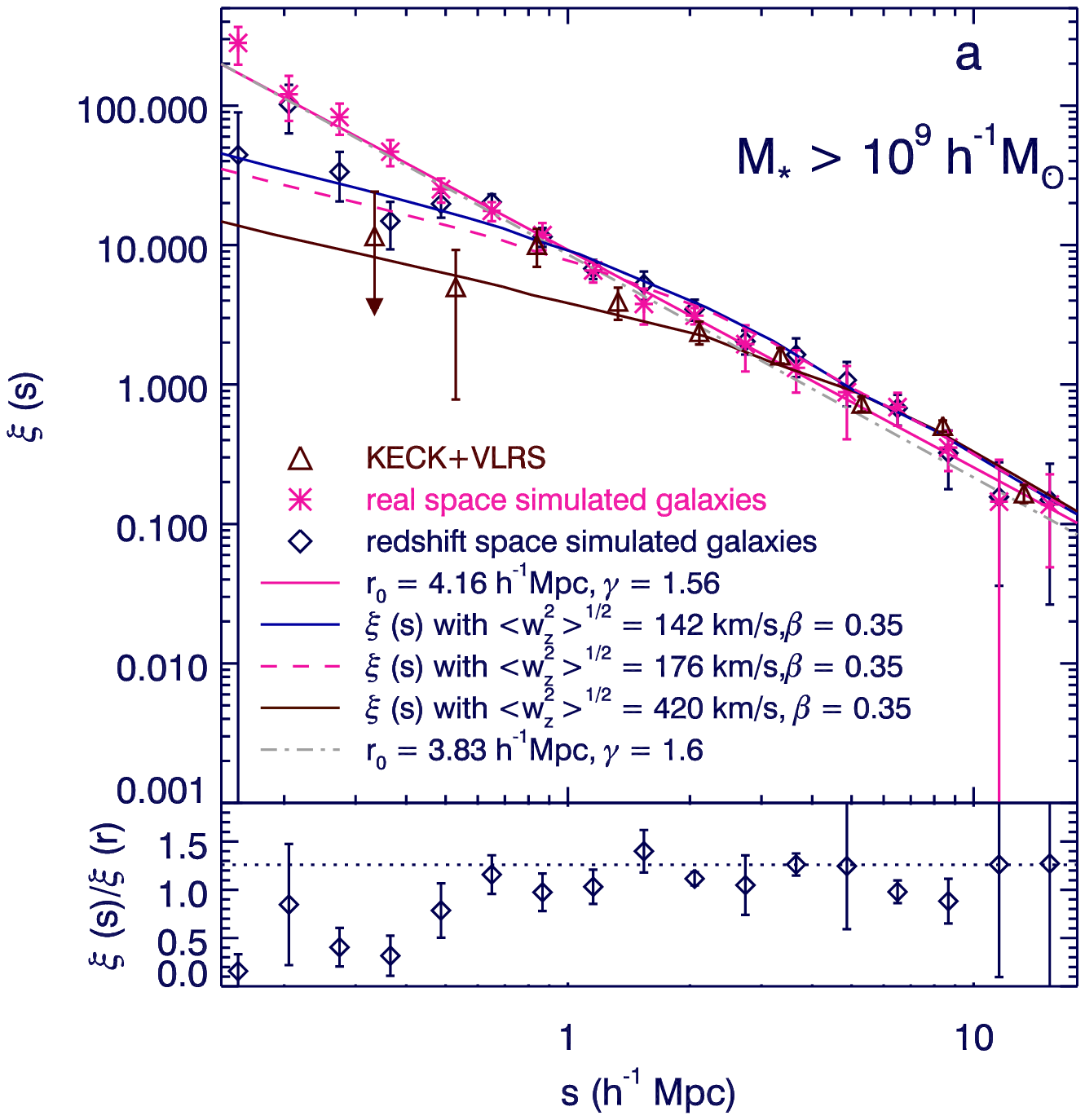}
\includegraphics[width=88.mm]{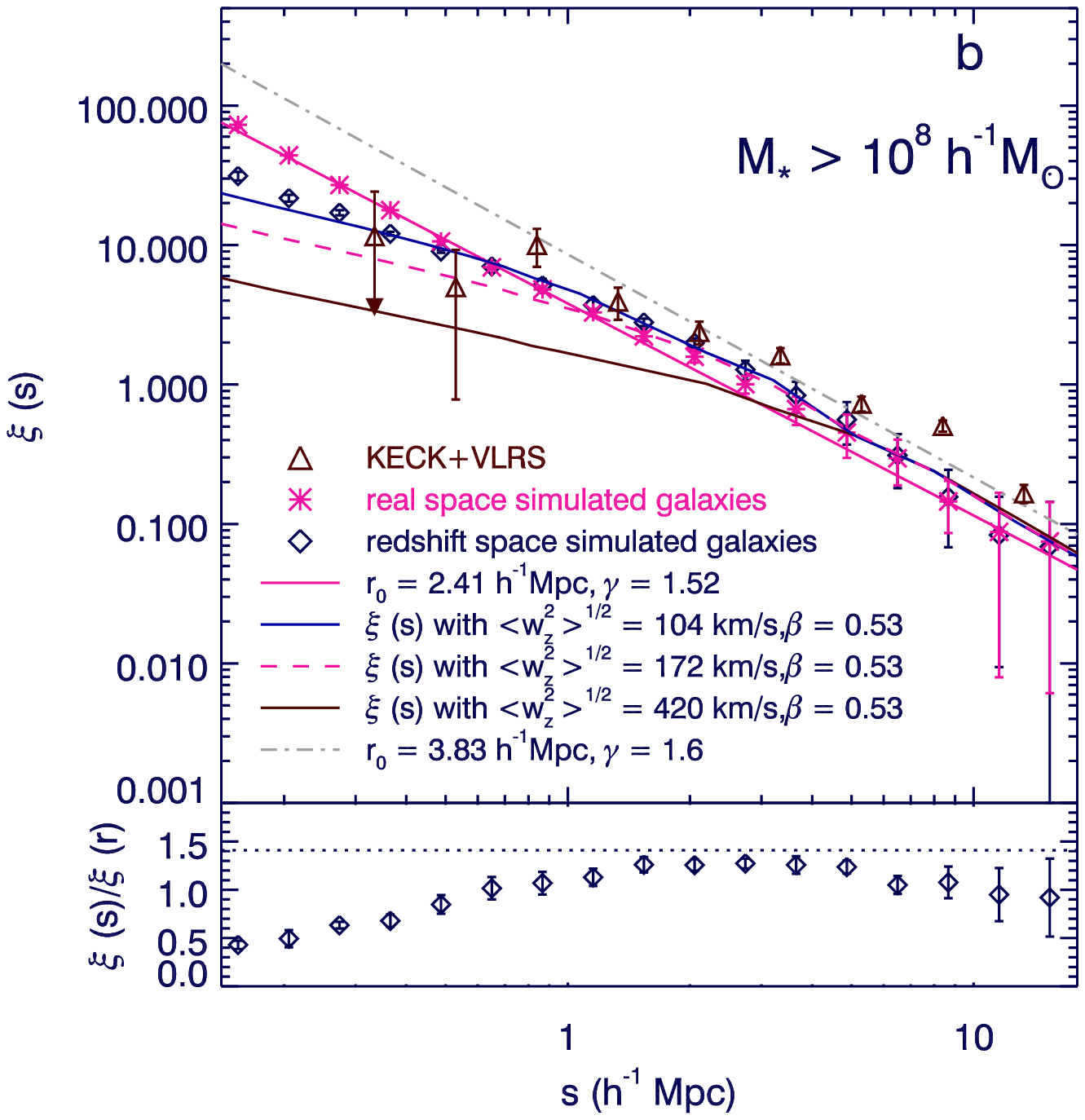} 
\caption{(a) Galaxy auto-correlation functions for 287 simulated \himass\
galaxies compared to the observed LBGs, with the GIMIC real-space result shown by pink asterisks,
the GIMIC redshift-space result shown by blue diamonds and the observed result of \citet{2013MNRAS.430..425B} given by brown triangles. The pink curve shows the power-law fit to the GIMIC real-space clustering, whilst the grey dot-dashed line shows the real space clustering derived from the observed sample. The pink dashed line is the predicted $\xi(s)$ assuming the real-space fit with $\vdisp$ = 176~km~s$^{-1}$ and $\beta=0.35$. The blue solid line is the same except with the $ r < 1 $~$h^{-1}$ Mpc pairwise dispersion of $\vdisp$ = 142~km~s$^{-1}$. The brown line is the RSD model with velocity errors added to allow comparison with the Keck+VLRS LBG $\xi(s)$. 
(b) The same for 4,070 simulated \lowmass\ galaxies with real-space fit $\gamma=\lgam$, $r_{0}=\lrzero~h^{-1}$Mpc. The $\xi(s)$ predictions now assume the appropriate pairwise velocity
dispersion of $\vdisp$ = 172~km~s$^{-1}$ (pink dashed line) and $\vdisp$ = 104~km~s$^{-1}$ (blue solid line). Bottom panels present $\xi(s)/\xi(r)$ with jack-knife error bars. The dotted line represents the predicted Kaiser boost with (a) $\beta_{\rm gal}=\hbeta$ giving $\xi(s)/\xi(r)=\hkaiser$ for \himass\ galaxies and the Kaiser boost with (b) $\beta_{\rm gal}=\lbeta$ giving $\xi(s)/\xi(r)=\lkaiser$ for \lowmass\ galaxies.}
\label{fig:xisLBGLBGsim} 
\end{figure*}

Fig.~\ref{fig:xisLBGLBGsim} shows the results for the simulated
galaxy-galaxy correlation function with (a) \himass\ and (b) \lowmass\
simulated galaxies. The blue diamonds show results from galaxies in
redshift-space while the pink asterisks show results from galaxies in
real-space. The integral constraint, ${\cal I}$, is included in the data
in order to compensate for the effect of the limited field sizes (as
described in \citealt{2013MNRAS.430..425B}). The estimated integral
constraints are ${\cal I}=\hintc$ and ${\cal I}=\lintc$ for \himass\ and the \lowmass\ galaxies respectively. The pink lines represent power-law fits to the real-space correlation function based on
Eq.~\ref{eq-power-law}. The power-law parameters for the fits to the clustering are given in Tab.~\ref{tab:galr0gam}. These power-law results give good fits to the
real-space clustering results and there is little sign of a
double power-law or two-halo break in the clustering for either of the samples. However, we note that in $z\sim3$ galaxies, the break between the 1-halo and 2-halo terms is measured to be at $\sim0.1'$ \citep{Hildebrandt2009}, which corresponds to $\approx0.14~\hmpc$ at $z=3$. Any break is therefore expected to be at scales smaller than those that we consider in Fig.~\ref{fig:xisLBGLBGsim}, scales at which we have little sensitivity with which to probe for any possible break.

\begin{table*}
\caption{Results for the power-law fits to the 1D galaxy auto-correlation functions.}
\centering
\begin{tabular}{lcccc}
\hline
Sample                           & $r_0$ ($\hmpc$) & $\gamma$      & Bias          & $\beta_{\rm gal}$       \\
\hline
GIMIC \lowmass                   & $2.41\pm0.24$   & $1.52\pm0.10$ & $1.85\pm0.12$ & $0.35\pm0.04$ \\
GIMIC \himass                    & $4.16\pm1.16$   & $1.56\pm0.26$ & $2.80\pm0.18$ & $0.23\pm0.08$ \\
VLRS \citep{2013MNRAS.430..425B} & $3.83\pm0.24$   & $1.60\pm0.09$ & $2.59\pm0.13$ & --- \\
\hline
\end{tabular}
\label{tab:galr0gam}
\end{table*}

\subsubsection{Simulated $\xi(s)/\xi(r)$ and infall}

In the lower panels of Fig.~\ref{fig:xisLBGLBGsim}, we show the ratio between the real and
redshift-space clustering results from the simulation in order to highlight the signatures
of RSD in the redshift-space correlation function. Here the errors are again constructed from the jack-knife realisations. At scales above $r\sim1.5-2~h^{-1}$Mpc, we see the effects
of dynamical infall, which acts to boost the clustering signal in the
redshift-space measurement by $\xi(s)/\xi(r)\sim1.2-1.4$. From linear theory (\citealt{kaiser1987,Hamilton1992}) we expect to see a `Kaiser boost' given given by:

\begin{equation}
\label{kaiser}
\xi(s)=\left(1+\frac{2}{\displaystyle 3}\beta_{\rm gal} + \frac{\displaystyle1}{\displaystyle5}\beta_{\rm gal}^2\right)\xi(r), 
\end{equation}

\noindent where $\beta_{\rm gal}$ is the dynamical infall parameter. For galaxies $\beta_{\rm gal}\approx\Omega^{0.6}/b$, where $b$ is the linear galaxy bias and is given by $b = \sqrt{\xi_{\rm{gal}}/\xi_{\rm{DM}}}$ (here $\xi_{\rm{gal}}$ is the galaxy clustering and $\xi_{\rm{DM}}$ is the dark matter clustering all in real-space, \citealt{kaiser1987}). At $z\approx 3$, we proceed via the volume averaged clustering amplitude, $\bar{\xi}(8)$, to evaluate both the galaxy and dark matter clustering and derive the bias - see Eqs 17, 18 of \citet{2013MNRAS.430..425B}.

Assuming the power-law fitted to $\xi(r)$ for the set of \lowmass\ galaxies, we find $\bar{\xi}_g(8)=\lxibar\pm\lxibare$, giving $b=\lbias\pm\lbiase$ and $\beta_{\rm gal}=\lbeta\pm\lbetae$. At separations of $1<r<8~h^{-1}\rm{Mpc}$, we find a mean amplitude ratio of $\lkairs\pm\lkairse$, which equates to an infall parameter of $0.35\pm0.04$. This is lower by $\approx2.5\sigma$ than the estimate based on the bias. For the \himass\ simulated galaxy case, the above power-law parameters fitted to $\xi(r)$ give $\bar{\xi}_g(8)=\hxibar\pm\hxibare$ which with $\bar{\xi}_{DM}(8)=0.088$ gives bias $b=\hbias\pm\hbiase$. Taking $\Omega_m(z = 3.0) = 0.98$ gives $\beta_{\rm gal}=\hbeta\pm\hbetae$. The measured Kaiser boost from $\xi(s)/\xi(r)$ is $\hkairs\pm\hkairse$, which equates to an infall parameter (based on Eq.~\ref{kaiser}) of $\beta_{\rm gal}=0.23\pm0.08$, consistent with what we would expect from the bias at the $\approx1.5\sigma$ level.

Overall, for both samples we find that the measurements based on the $\xi(s)/\xir(r)$ Kaiser boost appear to result in marginally lower values of $\beta$ than would be expected from the linear theory prediction based on $\beta=\Omega^{0.6}/b$, but only at a $\sim1-2\sigma$ level.

\subsubsection{Simulated galaxy correlations and velocity dispersion}

At smaller separations ($r<1$~$h^{-1}$ Mpc) for both high- and low-mass
simulated galaxies, the galaxy-galaxy $\xi(s)$ in redshift-space has
a lower amplitude than $\xi(r)$. This turn-over of the real-space
correlation function is the result of $z$-space smoothing due to the
pairwise velocity dispersion, $\vdisp$.  We model the effects of the
pairwise velocity dispersion on the clustering results using a Gaussian
profile to the velocity dispersion, following previous work
\citep[e.g.][]{Hawkins2003,DaAngela2005}:

\begin{equation}
f(w_z) = \frac{1}{\sqrt{2\pi}\vdisp}{\rm exp}\left(-0.5\frac{|w_z|^2}{\vdisp}\right)
\label{eq:vdisp}
\end{equation}

Using the pairwise velocity dispersions derived from Fig.~\ref{fig:gal-vz} (i.e.
$\vdisp=176~\rm{km~s}^{-1}$ and $\vdisp=172~\rm{km~s}^{-1}$ for the
\himass\ and \lowmass\ samples respectively - pink dashed lines in both
panels), we find that the reduction of the real-space clustering at
small scales is over-predicted compared to the measurements of $\xi(s)$.
As illustrated in Fig.~\ref{fig:gal-vz} however, we note that the
measured pairwise velocity dispersion is separation dependent. The
discrepancy is therefore likely the result of the effect of
small scale peculiar motions on the clustering function being dominated
by galaxies within $\sim1\hmpc$ of each other, whereas the initial
pairwise velocity histogram presented in Fig.~\ref{fig:gal-vz} includes
pairwise velocities  between galaxies across all separation scales
within the simulation. If we thus limit the histogram of pairwise
velocities to only those pairs within $1\hmpc$ of each other (dashed
histograms in Fig.~\ref{fig:gal-vz}), we retrieve pairwise velocity
dispersions of $\vdisp=142~\rm{km~s}^{-1}$ and $\vdisp$ = 104~km~s$^{-1}$
for the \himass\ and \lowmass\ samples respectively. Using these values
in the RSD model, we find improved agreement
between the model (solid blue line in Fig.~\ref{fig:xisLBGLBGsim}) and
the galaxy auto-correlation function measured from the GIMIC
simulations. Ultimately, the appropriate velocity dispersion for
modelling the RSD effects on the galaxy
clustering, is the velocity dispersion present within groups, whilst the
peculiar velocity measured from the simple histogram case included the
imprint of the velocity dispersion of galaxy groups as well as the
dispersion within groups. Taking the histogram of only pairs of galaxies
within $\sim 1\hmpc$ of each other effectively measures the intra-group
peculiar velocities. We conclude that $\xi(s)$ is better described on 
sub-Mpc scales with the intra-group velocity dispersion appropriate 
for these scales.

\subsection{Simulated and observed correlation functions compared}

\cite{2013MNRAS.430..425B} report  the best fit scale-length and  slope
for the observed Keck $+$ VLRS LBG-LBG semi-projected $w_p(\sigma)$ for
the data is $r_0 = 3.83 \pm 0.24~h^{-1}$ Mpc with a slope of $\gamma =
1.60 \pm 0.09$. Within the reported errors, the clustering of our
\himass\ sample reproduces the observed survey clustering very well in
terms of both clustering length and slope. As would be expected, the
\lowmass\ sample gives a somewhat lower clustering length than the
observational data, but does at least have a consistent slope within the
quoted errors.

We now apply the measured $\vdisp$ from the observations of
\citet{2013MNRAS.430..425B} to our correlation functions measure from
GIMIC. \citet{2013MNRAS.430..425B} measured $\vdisp=420~\rm{km~s}^{-1}$,
which includes both the intrinsic velocity dispersion and the velocity
errors on measuring the galaxy redshifts. The measured $\xi(s)$ from
\citet{2013MNRAS.430..425B} is shown in Fig.~\ref{fig:xisLBGLBGsim}
(brown triangles) and a model based on the GIMIC $\xi(r)$ combined with
the observational $\vdisp=420~\rm{km~s}^{-1}$ is given by the brown solid
line. By introducing the observationally measured pairwise velocity
errors to the GIMIC \himass\ result, we find that the GIMIC clustering
measurement reproduces well the measured LBG clustering.

\subsection{2-D correlation function}

We now turn to the 2-D galaxy auto-correlation functions in order
to further investigate the impact of galaxy velocities on clustering
measurements within the simulation. In the 2-D correlation function,
$\xi(\sigma,\pi)$, we parameterise the line of sight separation between
two galaxies by $\pi$ and the transverse separation by $\sigma$. We
calculate $\xi(\sigma,\pi)$ using the same methods as used for the 1-D
correlation functions and with the same samples.
 
\subsubsection{Simulations}

\begin{figure}
\centering
\includegraphics[width=86.mm]{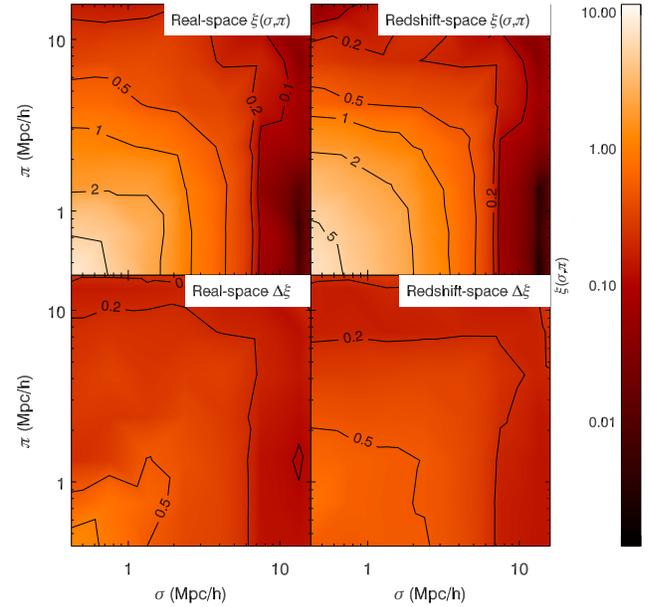}
\caption{The 2D auto-correlation function $\xi(\sigma,\pi)$ results based on the simulated \lowmass\ galaxies. The top panels show $\xi(\sigma,\pi)$ measured in real (left-panel) and redshift-space (right-panel), with a clear shift in the contours in the line-of-sight ($\pi$) direction at small scales showing the effect of peculiar velocities. Large scale bulk motions are also in evidence via the flattening of the $\xi=0.2$ contour at $\pi\sim10~h^{-1}$Mpc. The lower panels show the error contours over the same scales.} 
\label{fig:xisp_lbglbg_sim_lowmass}
\end{figure}

\begin{figure}
\includegraphics[width=86.mm]{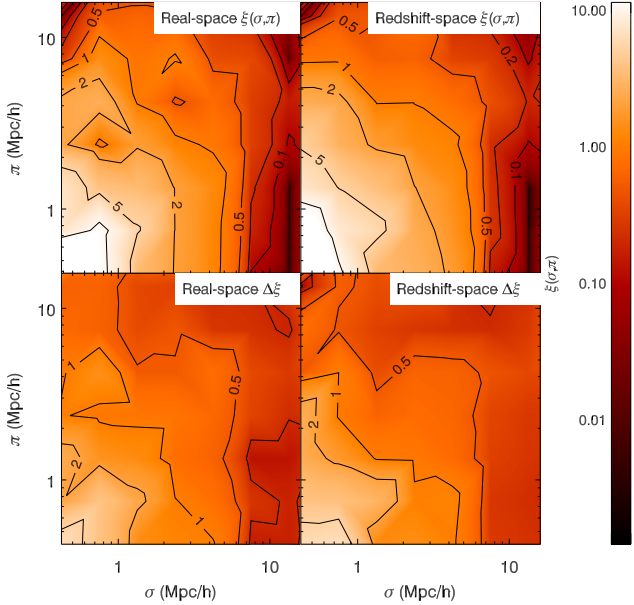}
\caption{As in Fig.~\ref{fig:xisp_lbglbg_sim_lowmass} but for the GIMIC \himass\ galaxy sample.} 
\label{fig:xisp_lbglbg_sim_himass}
\end{figure}

Figs.~\ref{fig:xisp_lbglbg_sim_lowmass} and \ref{fig:xisp_lbglbg_sim_himass} show the 2-D galaxy auto-correlation function, $\xi (\sigma,\pi)$, for \lowmass\ and \himass\ simulated galaxies respectively (both with the integral constraint added). In both cases the top-left panel shows the real-space measurement and the top-right panel shows the redshift-space measurement. The bottom panels show the respective error contours for the $\xi(\sigma,\pi)$ measurements.

Taking the \lowmass\ results first, the effects of the RSD are clearly visible in the top panels of Fig.~\ref{fig:xisp_lbglbg_sim_lowmass}, where the redshift-space $\xi(\sigma,\pi)$ contours are more extended at scales of $\lesssim4\hmpc$, whilst being flattened at scales of $\gtrsim4\hmpc$ in comparison to the real-space result. In terms of the latter, the shift in position of the $\xi=0.5$ and $\xi=0.2$ contours from the left to right panels is clear evidence of the Kaiser boost.

We now fit this $\xi(\sigma,\pi)$ result with a model based on incorporating the infall parameter, $\beta$, and convolving this with the velocity dispersion \citep[e.g.][]{Hawkins2003,DaAngela2005}:

\begin{equation}
\xi(\sigma,\pi)=\int_{-\infty}^{\infty}\xi^\prime(\sigma,\pi-w_z(1+z)/H(z))f(w_z){\rm d}w_z
\label{reddistf}
\end{equation}

\noindent where $\xi^\prime$ is given by:

\begin{equation}
\begin{split}
\xi^\prime(\sigma,\pi) = \left(1+\frac{2\beta_{\rm gal}}{3}+\frac{\beta_{\rm gal}^2}{5}\right)\xi_0(r)P_0(\mu) \\
+\left(4\frac{\beta_{\rm gal}}{3}+\frac{4\beta_{\rm gal}^2}{7}\right)\xi_2(r)P_2(\mu) \\
+ \frac{8\beta_{\rm gal}^2}{35}\xi_4(r)P_4(\mu)
\end{split}
\label{eq:xiprime_auto}
\end{equation} 

\noindent where $P_l(\mu)$ are Legendre polynomials, $\mu=cos(\theta)$ and $\theta$ is the
angle between $r$ and $\pi$. $\xi_0(r)$, $\xi_2(r)$ and $\xi_4(r)$ are the monopole, quadrupole and hexadecapole components of the linear $\xi(r)$. In general they are given by \citep{1996ApJ...470L...1M}:

\begin{equation}
\xi_{2l}(r) = \frac{-1^l}{r^{2l+1}}\left(\int^r_0x{\rm d}x\right)^l\left(\frac{\rm d}{{\rm d}x}\frac{1}{x}\right)^lx\xi(x)
\label{eq:poles}
\end{equation}

\begin{figure}
\centering
\includegraphics[width=86.mm]{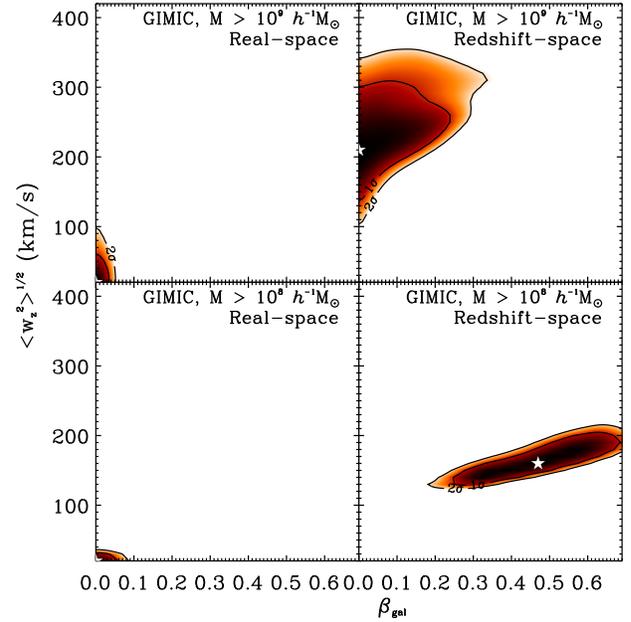}
\caption{The top panels show the RSD fitting results in real (left) and redshift-space (right) for the GIMIC $M_\star>10^9~h^{-1}{\rm M_{\odot}}$ galaxy sample. The real-space fitting is consistent with the lack of velocity effects in the data, giving best fitting parameters of $\vdisp=0^{+60}_{-0}\rm{km~s}^{-1}$ and $\beta_{\rm gal}=0.00^{+0.04}_{-0.00}$. In the redshift-space measurement, we find a velocity dispersion of $\vdisp=210^{+90}_{-70}\rm{km~s}^{-1}$. The large scale motions is constrained as $\beta_{\rm gal}=0.00^{+0.24}_{-0.00}$. The lower panels show the best fitting parameters to the real and redshift-space results using the RSD model described in the text for the GIMIC $M_\star>10^8~h^{-1}{\rm M_{\odot}}$ galaxy sample. Fitting to the real-space result gives parameters consistent with the null velocity field, with $\vdisp=0^{+30}_{-0}\rm{km~s}^{-1}$ and $\beta_{\rm gal}=0.00^{+0.06}_{-0.00}$ (left-panel). Applying the same model to the redshift-space $\xi(\sigma,\pi)$ we retrieve best fitting parameters of $\vdisp=160^{+45}_{-35}\rm{km~s}^{-1}$ and $\beta_{\rm gal}=0.47\pm0.22$ (right-panel), consistent with the simulated velocity field.} 
\label{fig:xisp_lbglbg_sim_betavdisp}
\end{figure}

Thw effect of RSDs is affirmed when fitting this RSD model as shown by the lower panels of Fig.~\ref{fig:xisp_lbglbg_sim_betavdisp}. The fitting is performed by applying the RSD model to the power-law fit given in Fig.~\ref{fig:xisLBGLBGsim}b (i.e. $r_0=2.41\hmpc$ and $\gamma=1.52$). We fit the model firstly to the real-space $\xi(\sigma,\pi)$ in order to constrain any geometric effects on the 2D clustering that may mimic RSD. The model fitting applied in real-space gives best fit parameters of $\vdisp=0^{+30}_{-0}\rm{km~s}^{-1}$ and $\beta_{\rm gal}=0.00^{+0.06}_{-0.00}$, consistent with this measurement having been made in real-space. Performing the same fitting to the redshift-space result returns best fit values of $\vdisp=160^{+45}_{-35}\rm{km~s}^{-1}$ and $\beta_{\rm gal}=0.47\pm0.22$. From the measured bias for the galaxy sample of $b=1.85$, we predicted an infall parameter value for this galaxy sample of $\beta_{\rm gal}=0.53\pm0.03$. Additionally, from the ratio of $\xi(s)/\xi(r)$, we find $\beta_{\rm gal}=0.35$, which again is within the $1\sigma$ errors of the 2D fitting result. As for the velocity dispersion, we find that the result is $>1\sigma$ higher than the result for the 1D clustering measurement ($\vdisp=104~{\rm km~s}^{-1}$), but is consistent with the intrinsic velocity dispersion measured from the galaxy sample directly ($\vdisp=172~{\rm km~s}^{-1}$).

Turning to the \himass\ galaxy sample, the top panels of Fig.~\ref{fig:xisp_lbglbg_sim_himass} show $\xi(\sigma,\pi)$ in real (left-panel) and redshift (right-panel) space (with the lower panels showing the error contours). The $\chi^2$ contours for the fits to the real and redshift-space measurements are shown in the top panels of Fig.~\ref{fig:xisp_lbglbg_sim_betavdisp}. The fitting was again made based on the $\xi(r)$ power-law fit (i.e. $r_0=\hrzero~h^{-1}$Mpc and $\gamma=\hgam$). The best fit for real-space is $\beta_{\rm gal}=0.00^{+0.04}_{-0.00}$ and velocity dispersion $\vdisp=0^{+60}_{-0}\rm{km~s}^{-1}$ with reduced $\chi^{2}$ = 0.7. In redshift-space, we found $\beta_{\rm gal}=0.00^{+0.24}_{-0.00}$ and $\vdisp=210^{+90}_{-70}\rm{km~s}^{-1}$ with reduced $\chi^2=0.7$.

\begin{table}
\caption{Results for the power-law fits to the 2D galaxy auto-correlation functions.}
\centering
\begin{tabular}{lcc}
\hline
Sample                           & $\beta_{\rm gal}$ & $\vdisp$ ($\kps$) \\
\hline
VLRS \citep{2013MNRAS.430..425B} & $0.38\pm0.19$          & $420_{-160}^{+140}$ \\
GIMIC \lowmass                   & $0.47\pm0.22$          & $160_{-35}^{+45}$ \\
GIMIC \himass                    & $0.00_{-0.00}^{+0.24}$ & $210_{-70}^{+90}$ \\
\hline
\end{tabular}
\label{tab:gal_2dfits}
\end{table}

The bias of $b=2.80$ suggests a value of $\beta_{\rm gal}\approx\Omega_m^{0.6}/b=\hbeta\pm\hbetae$, which is $>1\sigma$ different from the best fitting parameter given by the $\xi(\sigma,\pi)$ fitting. The fitted value of $\beta_{\rm gal}=0.00^{+0.24}_{-0.00}$ is however consistent at the $\approx1\sigma$ level with the $\beta_{\rm gal}=0.24$ implied by the ratio of $\xi(s)/\xi(r)$. In terms of the velocity dispersion fitting parameters, the 1D and 2D fitted $\vdisp$ values ($\vdisp=142~\rm{km~s}^{-1}$ and $\vdisp=210^{+90}_{-70}\rm{km~s}^{-1}$ respectively) are consistent at $\sim1\sigma$, although the 2D result is again higher than the $1\sigma$ result. These results are summarised in Table~\ref{tab:gal_2dfits}.

In summary, the analysis of $\xi(\sigma,\pi)$ from the simulation has shown that we may determine RSD effects using the 2D clustering consistently (at the $\sim1\sigma$ level) with the analysis of the 1D clustering. There is some tension for the \himass\ sample where the best fitting $\beta_{\rm gal}$ is zero, however this is still consistent with the 1D clustering analysis at the $1\sigma$ level. In all cases, the model successfully constrains the real-space clustering to be consistent with there being no RSD effects. In addition, the infall-parameter results are consistent with the linear theory analysis at the $1\sigma$ level in the case of the \lowmass\ sample and the $2\sigma$ level for the \himass\ sample.

Further to this, we have shown that the GIMIC galaxy population has consistent properties with observations of LBGs at $z\sim3$. For example, \cite{2013MNRAS.430..425B} presented the results for $\xi(\sigma,\pi)$ for $z\sim3$ LBGs, finding  $\beta(z=3)=0.38\pm0.19$, with $r_0 = 3.83\pm0.24$~$h^{-1}$ Mpc and $\gamma = 1.60\pm0.09$. The \himass\ galaxy clustering gives consistent values for all three of these parameters at the $1\sigma$ level. Unfortunately, the small scale velocity field for the observations is dominated by redshift errors, rather than the intrinsic galaxy peculiar velocities, so we have no suitable $z\sim3$ data to compare our small-scale results with. However, the results obtained from the simulation for $\vdisp$ are instructive for observational analyses.

\section{Galaxies and the IGM}
	
As discussed earlier, the relationship between the galaxy population and the IGM is key to understanding galaxy growth and evolution. Galaxies require large halos of gas in order to grow to the large masses we observe at the present day, whilst the supply and regulation of the flow of gas into galaxies dictates the distribution of galaxy masses we observe.

From observations of galaxy winds with speeds of $\gtrsim$~300~km~s$^{-1}$ for the LBG population (e.g. via the offsets nebulae and inter-stellar medium spectral features), it is evident that outflowing material exists in these star-forming galaxies \citep[e.g.][]{2001ApJ...554..981P,Shapley2003,Bielby2011}. A number of authors have thus attempted to detect the effects of such outflows on the distribution of gas around the $z\sim2-3$ star-forming galaxy population via the Ly$\alpha$ forest observed in the spectra of background sightlines (e.g. A03, A05, \citealt{Crighton2011,Rudie2012,Rakic2012}).

In this section, we perform an analysis of the cross-correlation between galaxies and the Ly$\alpha$ forest using both the VLRS observational data and the GIMIC simulation. We apply the same dynamical models as in the previous sections to the cross-correlation analysis. In the case of the galaxy-Ly$\alpha$ cross-correlation the relation between redshift and real-space correlations will become \citep{Mountrichas-Shanks2007}:

\begin{equation}
\xi(s)/\xi(r)=(1+\frac{1}{3}(\beta_{\rm gal}+\beta_{\rm Ly\alpha}) + \frac{1}{5}\beta_{\rm gal}\beta_{\rm Ly\alpha}),
\label{eq:kaiser-cross}
\end{equation}

The linear bias of the gas obtained from $b^2=\xi_{\rm Ly\alpha}/\xi_{\rm DM}$ is $b\approx0.3$ (see Section~\ref{sec:lya_auto}) but this is not the
bias required to assess the effect of gas infall via 
$\beta_{\rm Ly\alpha}$. This is because of the non-linear relation $F= e^{-\tau}$
between Ly$\alpha$ transmission and optical depth, $\tau$, where most
of the physics in the Ly$\alpha$ forest is contained in $\tau$. 
According to McDonald et al (2000, 2003) the
infall parameter $\beta_{\rm Ly\alpha}=\Omega_m^{0.6}\times b_\eta/b_\delta$ 
and $b_\eta$ and $b_\delta$ have to be determined from
simulations. \cite{McDonald2003} found results for
$\beta_{\rm Ly\alpha}=1-1.6$ depending on the resolution of the
simulations. We therefore take $\beta_{\rm Ly\alpha}=1.3$ as our estimate
of the gas dynamical infall parameter. \cite{McDonald2003}
did not use the RSD techniques used here so this and the
fact that we are using a higher resolution SPH simulation makes it
interesting to check whether linear theory with their $\beta_{\rm Ly\alpha}$
fits our simulated  data. \cite{McDonald2000} argue that the form of the
flux correlation function is proportional to the mass correlation
function in the linear regime. Following \cite{McDonald2003}, we shall assume that we can take account of `finger-of-God'
velocity dispersions in the usual way by convolving the transmission correlation
function with a Gaussian of the appropriate dispersion.

We perform the LBG-Ly$\alpha$ cross-correlation using the normalised pixel flux values along the quasar sightlines, where the normalised flux or transmissivity is given by:
	
	\begin{equation}
		T =  \frac{\bar T(z=3)}{\bar T(z)}\frac{f}{f_{con}},
 	\end{equation}
	
\noindent where $f$ is the observed flux at a given wavelength/Ly$\alpha$-redshift and $f_{con}$ is the flux continuum at that wavelength/Ly$\alpha$-redshift. Following A03, our derived values of $T$ incorporate a renormalisation to remove the redshift evolution from the normalised flux based on $\bar T$ which is given by:
	 	
	\begin{equation}
	\label{eq:tbarrenorm}
		\bar T(z)  = 0.676 - 0.220(z-3),
        \end{equation}
				
\noindent where $z$ is the redshift of a given pixel \citep{McDonald2000}. We do not include the forest at wavelengths below the intrinsic Ly$\beta$ emission of the quasars, in order to avoid regions contaminated by Ly$\beta$ absorption lines. Thus, only the spectrum between the Ly$\beta$ and Ly$\alpha$ is used in this calculation. We also excluded the wavelength range within 20 \AA\ of the intrinsic Ly$\alpha$ emission to avoid any proximity effects from the quasars.

We then use the transmissivity of the Ly$\alpha$ forest as calculated
above to perform the LBG-Ly$\alpha$ cross-correlation function.
The LBG-Ly$\alpha$ cross-correlation function is calculated from
			
	\begin{equation}
		\fltrans = \frac{\left<DT(s)\right>}{N(s)},
 	\end{equation}	

\noindent where $\left<DT(s)\right>$ is the number of galaxy-Ly$\alpha$ pairs
weighted by the normalised transmissivity for each separation.
$N(s)$ is the number of LBGs that contribute to the cross-correlation
function at each separation.
	
\subsection{Observed LBG-Ly$\alpha$ cross-correlation}

\begin{figure*}
	\centering
\includegraphics[width=180.mm]{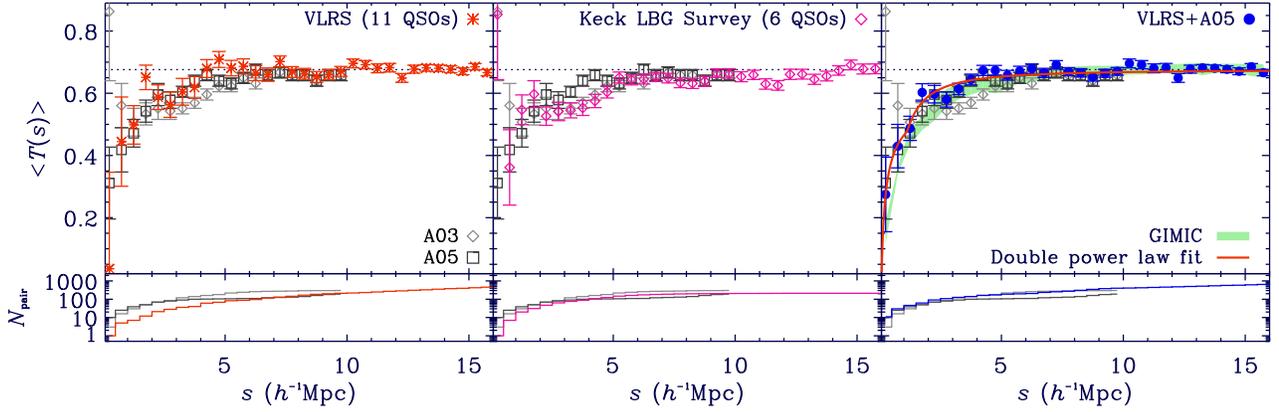}
	\caption{The mean Ly$\alpha$ transmissivity as a function of distance, $s$, from galaxies in observed $z\approx3$ samples. The left hand panel shows the result for the 11 sightlines observed as part of the VLRS alone; the central panel shows the result for the 6 sightlines observed with Keck; and the right hand panel shows the result for the VLRS result combined with the result of A05. The lower panels in each case show the number of galaxy-sightline pairs within a given separation. In each of the panels, we also show the results of A03 (grey diamonds) and A05 (grey squares) for comparison. In the right hand panel, we also show the result of a double power-law fit to the VLRS$+$Keck data (red curve), the parameters for which are given in Tab.~\ref{tab:lbglyar0gam}. We also show in the right hand panel the $\fltrans$ result from the GIMIC analysis given by the green shaded region.}
	\label{fig:xcor_lyalbg_data}
	\end{figure*}

\begin{table*}
\caption{Results for the power-law fits to the 1D galaxy-Ly$\alpha$ cross-correlation functions.}
\centering
\begin{tabular}{lccccc}
\hline
Sample                           & $s_{0,s}$ ($\hmpc$) & $\gamma_s$    & $s_{0,l}$ ($\hmpc$) & $\gamma_l$    & $\beta_{Ly\alpha}$ \\
\hline
VLRS - from $\xi(s)$             & $0.08\pm0.04$       & $0.47\pm0.10$ & $0.49\pm0.32$ & $1.47\pm0.91$ & --- \\
\hline
Sample                           & $r_{0,s}$ ($\hmpc$) & $\gamma_s$    & $r_{0,l}$ ($\hmpc$) & $\gamma_l$    & $\beta_{Ly\alpha}$ \\
\hline
VLRS - from $w_p(\sigma)$       & $0.020_{-0.018}^{+0.074}$& $0.37_{-0.14}^{+0.45}$ & $0.59_{-0.20}^{+0.90}$     & $1.10\pm0.74$ & --- \\
GIMIC \lowmass                   & $0.10\pm0.07$       & $0.46\pm0.22$ & $0.51\pm0.39$     & $1.25\pm0.61$ & $0.27\pm0.05$ \\
GIMIC \himass                    & $0.16\pm0.09$       & $0.46\pm0.19$ & $0.61\pm0.34$     & $1.18\pm0.43$ & $0.31\pm0.07$ \\
\hline
\end{tabular}
\label{tab:lbglyar0gam}
\end{table*}

\subsubsection{1D cross-correlation, $\fltrans$}
	
In Fig.~\ref{fig:xcor_lyalbg_data}, we present the latest result for the LBG-Ly$\alpha$ cross-correlation from the VLRS (left panel: red asterisks). This covers a broad range of scales, measuring to separations of $s\approx20\hmpc$. Errors on the data-points are calculated by taking the standard deviation of the $\fltrans$ measure across all the individual galaxies contributing to a given bin, divided by the square root of the number of galaxies contributing to that bin.  We see an overall continuous decrease in Ly$\alpha$ transmission down to the minimum scale probed of $s=0.25\hmpc$ (although this smallest bin contains only a single galaxy).

We also show the LBG-Ly$\alpha$ transmissivity correlation function for the publicly available Keck data that we incorporate into our 2D analysis (centre panel: pink diamonds); and the A05 result combined with our own VLRS result (right panel: blue circles). In each panel, we also show the results of A03 (grey triangles) and A05 (grey squares). We note in passing that our own reductions of the Keck sample HIRES data gives results consistent with A03 LBG-Ly$\alpha$ results. At separations below $s\approx5~h^{-1}$ Mpc, the combined sample has the same trend as A05, with no evidence for a turn-up at $s<1$~$h^{-1}$Mpc, a feature that was claimed by A03 to be evidence for feedback. With the larger sample of LBGs close to quasar sightlines compared to \citet{Crighton2011}, we have now strengthened the evidence against feedback strongly decreasing Ly$\alpha$ absorption on $s\lesssim1\hmpc$ scales around galaxies.

\subsubsection{2-D cross-correlation, $\xi(\sigma,\pi)$}

\begin{figure*}
\centering
\includegraphics[height=60.mm]{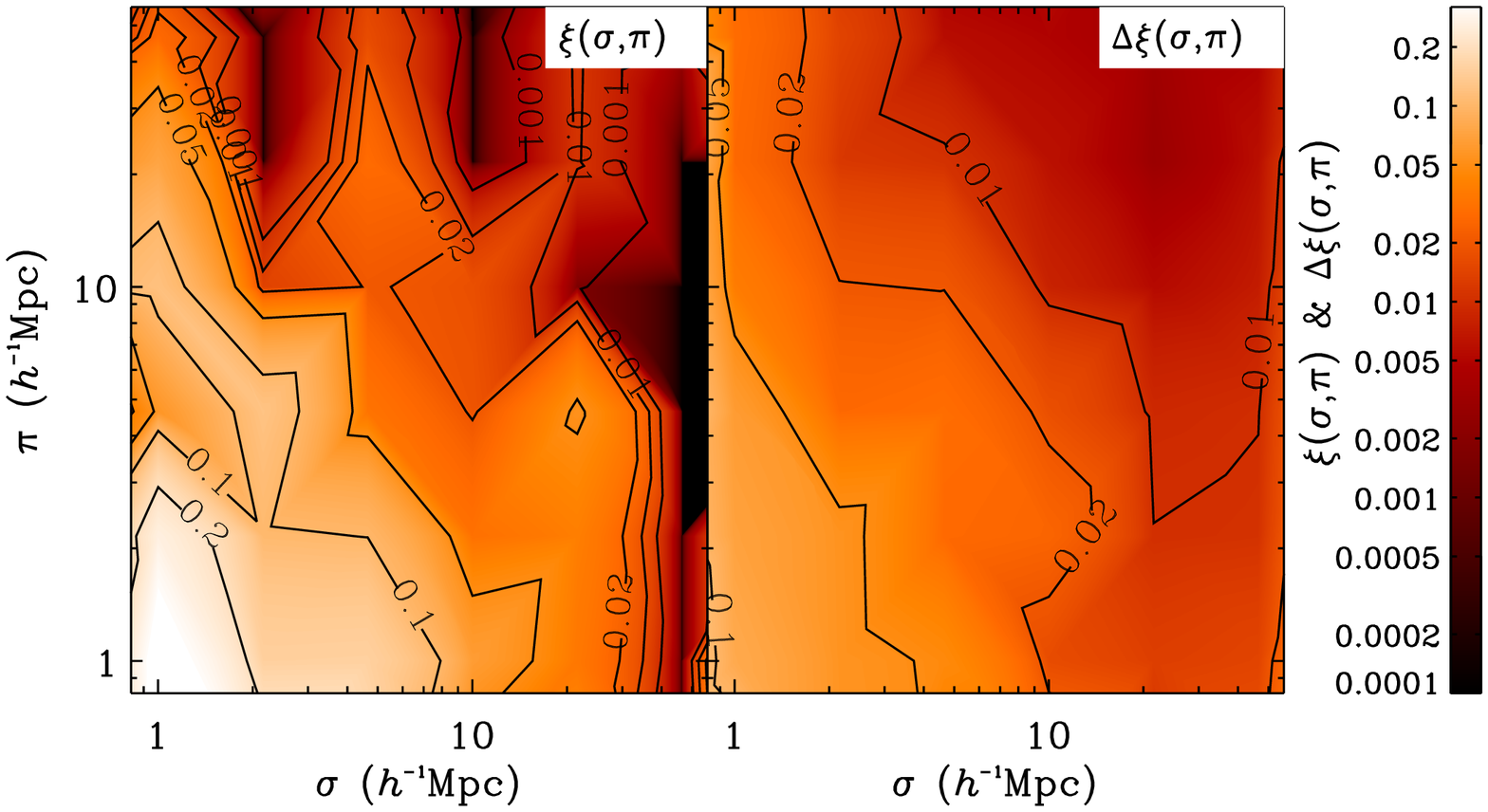}
\hspace{0.5cm}
\includegraphics[height=60.mm]{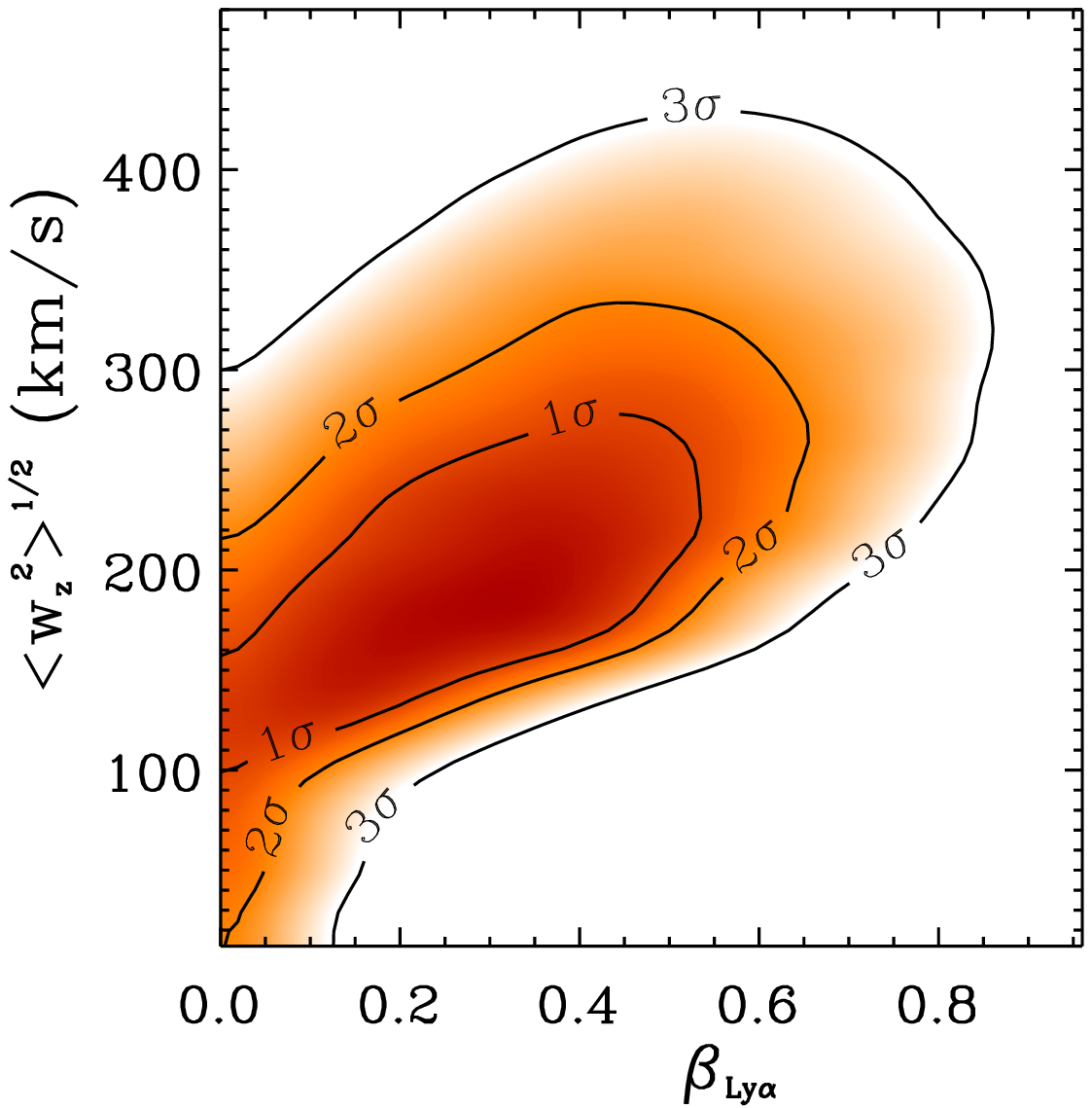}
\caption{The left hand panels show the LBG-Ly$\alpha$ $\xi (\sigma,\pi)$ and jack-knife errors on $\xi (\sigma,\pi)$ for the combined Keck$+$VLRS data. The right hand panel shows the result of fitting the $\xi(\sigma,\pi)$ model to the data, with best fit parameters given by $\beta_{\rm Ly\alpha}=0.33^{+0.23}_{-0.33}$ and $\vdisp=190\pm90~\rm{km~s}^{-1}$ (assuming an underlying double power-law form as given in Tab.~\ref{tab:lbglyar0gam} and with $\beta_{\rm gal}$ = 0.38).}
\label{fig:xisplyaLBG}
\end{figure*}

We now use the latest VLRS data sample of $\approx2,000$ LBGs alongside the Keck-based LBG-Ly$\alpha$ dataset to measure the 2-D LBG-Ly$\alpha$ cross-correlation, $\xi(\sigma,\pi)$. By combining these two surveys, we can compare the correlation functions in a wider range of separations than would otherwise be possible (the VLRS giving $~2-3\times$ the coverage in the $\sigma$ scale compared to the Keck data). The LBG-Ly$\alpha$ $\xi (\sigma,\pi)$ from Keck$+$VLRS sample is presented in Fig.~\ref{fig:xisplyaLBG}.

In order to fit the RSD model to this data, we first need an estimate of the real-space auto-correlation function. The double power-law fit to $\fltrans$ is unsuitable as it contains within it the imprint of the RSD effects. We therefore follow the usual route to estimating the real-space clustering and calculate the projected correlation function, $w_p(\sigma)$. This is calculated by integrating the 2D correlation function along the line of sight direction, $\pi$:

\begin{equation}
w_p(\sigma)=2\int^\infty_0\xi(\sigma,\pi){\rm d}\pi
\end{equation}

\begin{figure}
\centering
\includegraphics[height=60.mm]{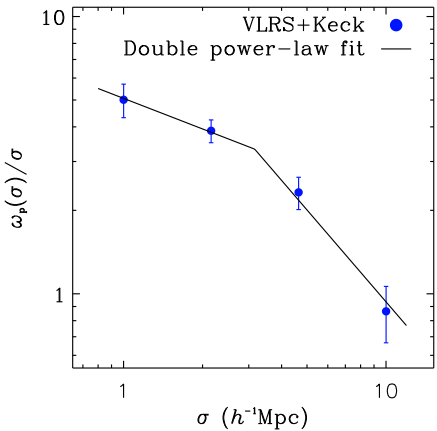}
\caption{The observed LBG-Ly$\alpha$ $w_p(\sigma)/\sigma$ result based on the VLRS$+$Keck galaxy and quasar sightline dataset (filled blue cirlces). A double power law fit is shown, the parameters for which are given in Tab.~\ref{tab:lbglyar0gam}.}
\label{fig:wpsig}
\end{figure}

The result is shown in Fig.~\ref{fig:wpsig}. At small scales ($\sigma\lesssim2\hmpc$), the clustering measurement will have a flatter slope due to the saturation of Ly$\alpha$ lines in the forest and so we fit the $w_p(\sigma)$ measurement with a double power-law. This is only marginally necessary given the error estimates on the measured $w_p(\sigma)$ data-points and is in part motivated by the analysis of the simulated sightlines that follows (see Fig.~\ref{fig:xis_lyalbg}). Each power-law takes the form:

\begin{equation}
\frac{w_p(\sigma)}{\sigma}=C\xi(\sigma) = C\left(\frac{r_0}{\sigma}\right)^\gamma
\end{equation}

\noindent where $C$ is given by:

\begin{equation}
C = \frac{\Gamma\left(\frac{1}{2}\right)\Gamma\left(\frac{\gamma-1}{2}\right)}{\Gamma\left(\frac{\gamma}{2}\right)}
\end{equation}

The resulting best fit parameters assuming this double power-law are given in Tab.~\ref{tab:lbglyar0gam}. We then use this fit as the basis with which to fit for RSD in the $\xi(\sigma,\pi)$ measurement. As in the galaxy-galaxy auto-correlation analysis, we use a model incorporating a Gaussian form for the effects of pairwise velocities, characterised by $\vdisp$, but now with the large scale infall characterised by a combination of $\beta_{\rm Ly\alpha}$ and $\beta_{\rm gal}$ (where $\beta_{\rm gal}$ is constrained by the auto-correlation results). The model is identical to that described earlier, except Eq.~\ref{eq:xiprime_auto} is now replaced by:

\begin{equation}
\begin{split}
\xi^\prime(\sigma,\pi) = \left(1+\frac{\beta_{\rm gal}+\beta_{\rm Ly\alpha}}{3}+\frac{\beta_{\rm gal}\beta_{\rm Ly\alpha}}{5}\right)\xi_0(r)P_0(\mu) \\
+\left(2\frac{\beta_{\rm gal}+\beta_{\rm Ly\alpha}}{3}+\frac{4\beta_{\rm gal}\beta_{\rm Ly\alpha}}{7}\right)\xi_2(r)P_2(\mu) \\
+ \frac{8\beta_{\rm gal}\beta_{\rm Ly\alpha}}{35}\xi_4(r)P_4(\mu)
\end{split}
\label{eq:xiprime}
\end{equation} 

\noindent where $P_l(\mu)$ are again the Legendre polynomials, $\mu=cos(\theta)$ and $\theta$ is the
angle between $r$ and $\pi$. $\xi_0(r)$, $\xi_2(r)$ and $\xi_4(r)$ are the monopole, quadrupole and hexadecapole components of the linear $\xi(r)$ and are given in Eq.~\ref{eq:poles}.

The resulting $\Delta \chi^2$ contours for this fit are shown in the right hand panel of Fig.~\ref{fig:xisplyaLBG}, with the best fitting result given by $\beta_{\rm Ly\alpha}=0.33^{+0.23}_{-0.33}$ and $\vdisp=190\pm90~{\rm km~s}^{-1}$ (given $\beta_{\rm gal}=0.38$).

As discussed, \citet{McDonald2003} predict a value for the infall parameter for the Ly$\alpha$ forest at $z=3$ of $\beta_{\rm Ly\alpha}=1.3\pm0.3$. Our measured value of $\beta_{\rm Ly\alpha}=0.33$ is more than $3\sigma$ lower than this predicted value. Comparing to other observations, \citet{Slosar2011} report a range of $0.44<\beta_{\rm Ly\alpha}<1.20$, at central redshift $z = 2.25$, from the analysis of BOSS quasar spectra. This is consistent at the $1\sigma$ level with our result, although at a lower redshift.

Predicting the velocity dispersion, we take the pairwise velocity dispersion measured for the galaxies ($\vdisp=420\kps$ - \citealt{2013MNRAS.430..425B}), which includes both the intrinsic dispersion and the velocity measurement errors, and combine this with the predicted velocity dispersion measured from the GIMIC simulation earlier ($120\kps$). As the galaxy measurement is a `pairwise' velocity, we thus need to divide this by $\sqrt{2}$, and therefore would expect $\vdisp=\sqrt{297^2+120^2}=320~\kps$ for the galaxy-Ly$\alpha$ $\vdisp$. The result obtained from the LBG-Ly$\alpha$ $\xi(\sigma,\pi)$ is consistent with this predicted value within the $2\sigma$ contours. We shall return to these Keck$+$VLRS results to compare with the results from the GIMIC simulations described below.

This measurement of the 2D LBG-Ly$\alpha$ cross-correlation is one of only a few such measurements, and the only one to give a full parameterised model fitting to the RSD. \citet{Rakic2012} and \citet{2014arXiv1403.0942T} show the 2D LBG-H{\sc i} pixel-optical-depth (POD) cross-correlation, giving estimated velocity dispersions of $\vdisp\sim240$~km~s$^{-1}$ and $\vdisp\sim260$~km~s$^{-1}$ respectively. There are significant differences between our analysis and these two authors, not least that they analyse a broader range in optical depth by including higher order Lyman series lines, but we note that our measured velocity dispersion is consistent with their results.

\subsection{LBG-Ly$\alpha$ cross-correlation from simulations}

As with the data, we compute the LBG-Ly$\alpha$ cross-correlation using the methods described above. We note however that the renormalisation to $z=3$, given by Eq.~\ref{eq:tbarrenorm}, is here redundant given that the simulated gas and galaxies are all at the same epoch already.

\subsubsection{Coherent motion of gas and galaxies}
\label{comotion}

\begin{figure*}
\centering
\includegraphics[width=120.mm]{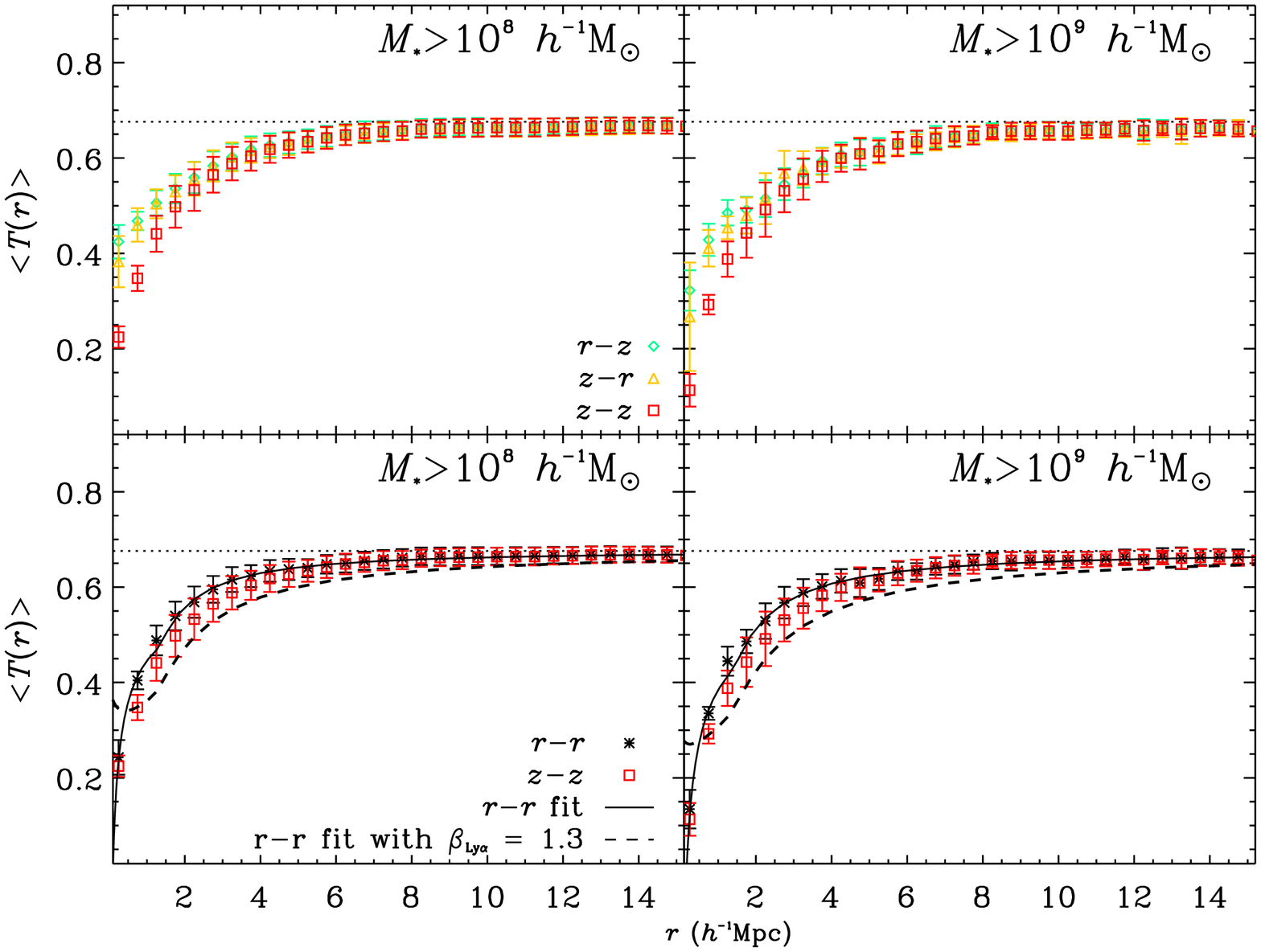}	
\caption{The transmissivity profile, $\fltranr$, around simulated galaxies within GIMIC for the \lowmass\ (left-hand panels)	and \himass\ (right-hand panels) galaxy samples. The top panels show the cross-correlation calculated using combinations of the galaxies in real space with the gas in redshift space (green diamonds), the galaxies in redshift space with the gas in real space (yellow triangles) and both galaxies and gas in redshift space (red squares). The lower panels show the same redshift space galaxy-Ly$\alpha$ cross-correlation (red squares) compared to the real-space cross-correlation (black asterisks). For both the \lowmass\ (left) and \himass\ (right) results we show a double power-law fit to the real-space result (blue line) and this fit convolved with the predicted RSDs (i.e. $\beta_{\rm Ly\alpha}=1.3$ with $\vdisp=139~\kps$ and $\beta_{\rm gal}=0.53$ for the \lowmass\ sample; and $\vdisp=156~\kps$ and $\beta_{\rm gal}=0.35$ for the \himass\ sample). All error bars were calculated using the jack-knife method.}
\label{fig:trans_lyalbg}
\end{figure*}

In the top panels of Fig.~\ref{fig:trans_lyalbg}, we show the Ly$\alpha$ mean transmissivity as a function of sightline-galaxy separation for the \lowmass\ (left-panel) and \himass\ (right-panel) galaxy samples and for three combinations of the gas and galaxies from the GIMIC simulation: galaxies in real space with the Ly$\alpha$ in redshift space ($r-z$, green diamonds); galaxies in redshift space with the Ly$\alpha$ in real space ($z-r$, yellow triangles); and both the galaxies and Ly$\alpha$ in redshift space ($z-z$, red squares). As in previous plots of $\fltrans$, the results are scaled to the mean transmissivity at $z=3$ (i.e. $\bar T(z=3)=0.676$). It is interesting to note that the decrease to smaller scales is enhanced as we go from the $r-z$ (or $z-r$) combination to the $z-z$ combination. If we assume that random Gaussian motions dominate galaxy peculiar motions, then this is a surprising result. The same effect is seen for both the \lowmass\ and \himass\ galaxy samples. This is however simply the result of a large ($\sim100~\kps$) bulk flow of material within the simulation volume - i.e. the analysis has not been performed at the mean rest frame of the particles in the box. Indeed this bulk motion is clearly evident in Fig.~\ref{fig:galRZ}.

The lower panels of Fig.~\ref{fig:trans_lyalbg} again show the galaxy-Ly$\alpha$ mean transmissivity as a function of sightline-galaxy separation for the \lowmass\ (left-panel) and \himass\ (right-panel) galaxy samples, but this time for the combinations of both galaxies and Ly$\alpha$ in real space ($r-r$, blue asterisks); and both the galaxies and Ly$\alpha$ in redshift space ($z-z$, red squares).

Focussing on the \lowmass\ galaxies, we see that both the $r-r$ and $z-z$ results show the same trends. At a distance $r>5~h^{-1}$Mpc, the measured $\fltranr$ increases towards the mean value.  As separations decrease below $5~h^{-1}$Mpc, the transmissivity decreases indicating an increase in the H{\sc{i}} density as we approach the galaxy. In terms of the effects of RSD, at separations of $r\sim1-6~h^{-1}$ Mpc we see that the galaxy-Ly$\alpha$ transmissivity correlation function in redshift-space lies lower than the real-space cross-correlation function. This behaviour is suggestive of the impact of coherent infall on the measured $z-z$ cross-correlation function.

To further investigate this, we perform a fit to the correlation function using a power-law form given by $\fltranr = \left(1-(r_0/r)^\gamma\right)\bar T(z=3)$. We first attempted a single power-law fit, but found that this failed to match both the large and small scale trends. This is primarily due to the non-linear nature of the relationship between the normalised flux measurement and the gas density, whereby, given a high enough column density of neutral hydrogen, the absorption line will reach zero flux and saturate. At this point, the normalised flux no longer gives a measure of increasing gas density and simply asymptotes to a value of zero. This has a significant effect on our measure of $\fltranr$, where close to galaxies the increasing mean gas density leads our measure of $\fltranr$ to turn over. We approximate this behaviour with a double power-law function: one power-law fitted to the large scale trend (i.e. $r\gtrsim1.6~h^{-1}{\rm Mpc}$); and another to approximate the small scale curtailing of $\fltranr$. For the \lowmass\ sample, we find best fitting parameters of $r_{0,s}=0.10\pm0.07\hmpc$, $\gamma_{s}=0.46\pm0.22$, $r_{0,l}=0.51\pm0.39\hmpc$ and $\gamma_{l}=1.25\pm0.61$ (where subscript $s$ denotes the small scale power-law and subscript $l$ denotes the large scale power law parameters). This fit is plotted as the solid black curve in the lower-left panel of Fig.~\ref{fig:trans_lyalbg} (and is summarised in Tab.~\ref{tab:lbglyar0gam}).

As a first step in analysing the RSD effects on the cross-correlation, we transform this fitted real-space fit to the GIMIC \lowmass\ sample cross-correlation with our RSD model and some reasonable estimates of what we may expect the RSD parameters to be. For the galaxy coherent large-scale motion, we have a value of $\beta_{\rm gal}=0.53$ derived from the galaxy-galaxy auto-correlation. For the Ly$\alpha$ coherent large-scale motion, we take $\beta_{\rm Ly\alpha}=1.3$ as predicted by the simulations of \citet{McDonald2003}. Finally, for the velocity dispersion parameter, we take $\vdisp=\sqrt{\left(104/\sqrt{2}\right)^2+120^2}=139~\kps$, i.e. combining the measured galaxy velocity dispersion (from Fig.~\ref{fig:gal-vz}) and Ly$\alpha$ velocity dispersion (from Fig.~\ref{fig:lya_vel}) in quadrature. The result is given by the black dashed line in the lower-left panel of Fig.~\ref{fig:trans_lyalbg}.

We perform an identical analysis with the \himass\ sample, fitting a double power law to the real-space $\fltranr$, finding best fit parameters of $r_{0,s}=0.16\pm0.09\hmpc$, $\gamma_{s}=0.46\pm0.19$, $r_{0,l}=0.61\pm0.34\hmpc$ and $\gamma_{l}=1.18\pm0.43$ (shown by the solid black curve in the  lower-right panel of Fig.~\ref{fig:trans_lyalbg}). The RSD model based on this double power-law fit is shown by the dashed black line in the  lower-right panel of Fig.~\ref{fig:trans_lyalbg} and is based on parameter values of $\beta_{\rm gal}=0.35$, $\beta_{\rm Ly\alpha}=1.3$ and $\vdisp=\sqrt{\left(142/\sqrt{2}\right)^2+120^2}=156~\kps$.

It is evident that the selected parameters do not provide a good fit to the redshift-space results from the GIMIC simulation in either case. For both the \lowmass\ and \himass\ samples, the model over predicts the effects of the coherent infall (at scales of $\gtrsim2\hmpc$) and the velocity dispersion at smaller scales. We investigate this further in the following sections.

\subsubsection{Dynamical Infall in $\xi(r)$}

\begin{figure*}
\centering
\includegraphics[width=120.mm]{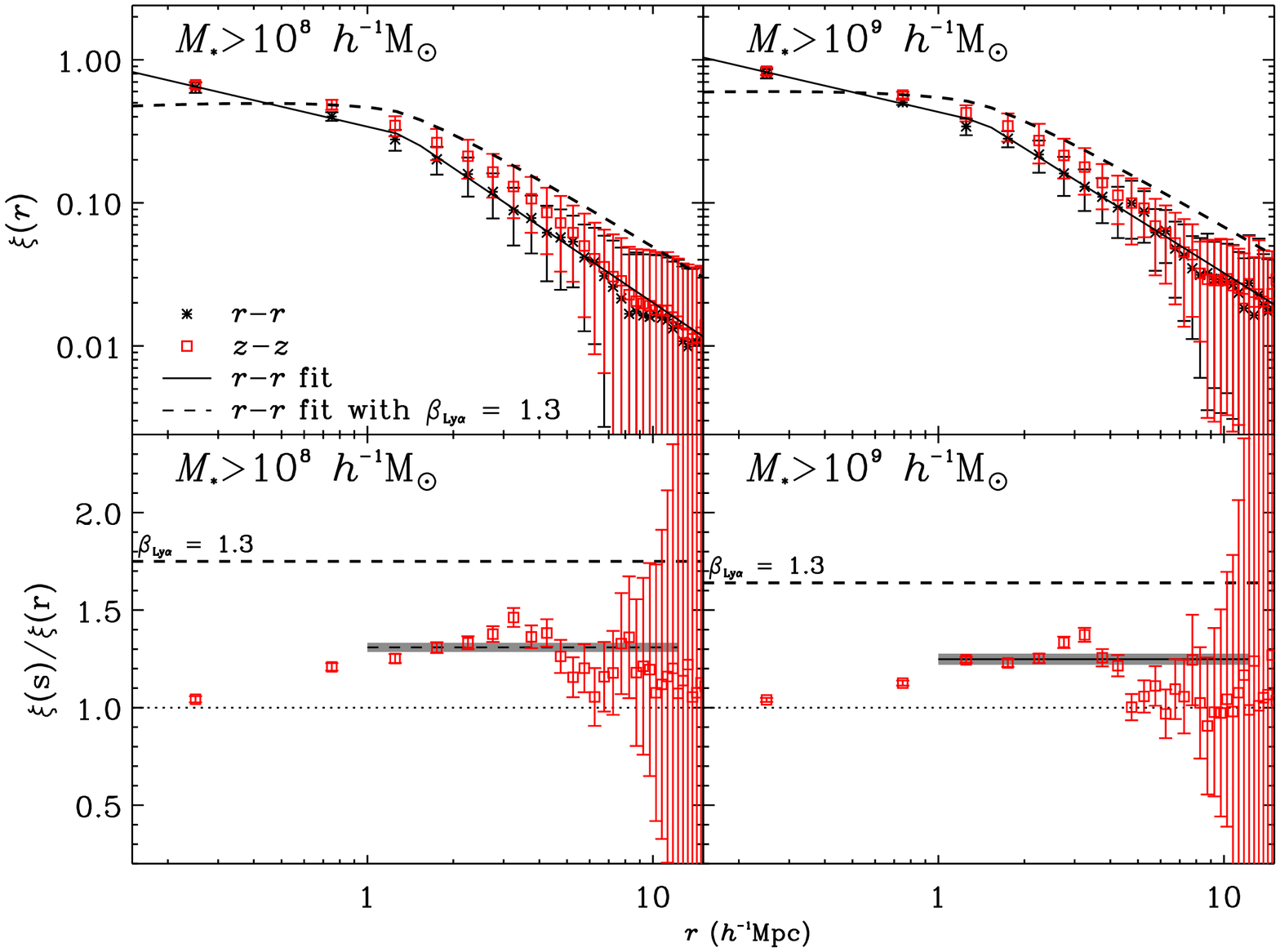}
\caption{The top panels show the LBG-Ly$\alpha$ cross-correlation function, $\xi(r)$, as derived from the $\fltranr$ profiles shown in Fig.~\ref{fig:trans_lyalbg}. Results for the \lowmass\ sample are shown in the left hand panels and the \himass\ sample in the right hand panels. In each case, we show the same fits as shown in Fig.~\ref{fig:trans_lyalbg} (solid black curves) and the subsequent redshift space distorted predictions based on a value of $\beta_{\rm Ly\alpha}= 1.3$ (dashed black curves). The lower panels show the ratio of the redshift-space cross-correlation functions, $\xi(s)$, to the real-space cross-correlation functions, $\xi(r)$. The dashed black lines in both lower panels shows the large scale prediction for $\xi(s)/\xi(r)$ assuming $\beta_{\rm Ly\alpha}=1.3$. The solid black lines and grey regions show the weighted mean of $\xi(s)/\xi(r)$ measured at $r>1\hmpc$ and the $1\sigma$ errors on the weighted mean.}
\label{fig:xis_lyalbg}
\end{figure*}

To better visualise any distortions in the cross-correlation, we calculate the function $\xi(r) = 1 - {\fltranr/\bar T(z=3)}$. The results for $\xi(r)$ are shown in the top panels of Fig.~\ref{fig:xis_lyalbg}. The points and curves are the same as given in the lower panels of Fig.~\ref{fig:trans_lyalbg} (except transformed from $\fltranr$ to $\xi(r)$) and again the \lowmass\ and \himass\ samples are shown in the left and right hand panels respectively. The models used for the curves are identical to those given in the previous section, but we now see more clearly why a single power law is unable to provide a good fit to the real-space data points in both the \lowmass\ and \himass\ cases. It is also clearer in these plots how the $\beta_{\rm Ly\alpha}$, $\vdisp=139~\kps$/$\vdisp=156~\kps$ RSD models provide a poor fit to the redshift space $\xi(s)$ results (red squares). The model lies at $\sim1\sigma$ above the data at all points above $\sim1\hmpc$, whilst it also over predicts the effects of the small scale velocity dispersion. This is the case for both the \lowmass\ and \himass\ samples.

It is the $\beta_{\rm Ly\alpha}=1.3$ value that is proving too high here, resulting in the model tending to over-predict the galaxy-Ly$\alpha$ cross-correlation function. In the lower panels of Fig.~\ref{fig:xis_lyalbg}, we show the ratio $\xi(s)/\xi(r)$. We measure a weighted average of the ratio over scales of $1\geq r\geq12\hmpc$ of $\left<\xi(s)/\xi(r)\right>=1.29\pm0.02$ and $\left<\xi(s)/\xi(r)\right>=1.24\pm0.03$ for the \lowmass\ and \himass\ samples respectively. Via Eq.~\ref{kaiser}, these values correspond to $\beta_{\rm Ly\alpha}=0.27\pm0.05$ (\lowmass) and $\beta_{\rm Ly\alpha}=0.31\pm0.07$ (\himass). 

From the $\xi(s)$ measurement, we are thus able to place constraints on a measure of the infall of gas towards galaxies, via the $\beta_{\rm Ly\alpha}$ quantity, consistently obtaining $\beta_{\rm Ly\alpha}\sim0.3$ for both of our galaxy samples. We now move to the 2-D cross-correlation function to evaluate if we obtain consistent results with a 2-D analysis.

\subsubsection{Dynamical Infall in $\xi (\sigma, \pi)$}

\begin{figure}
\centering
\includegraphics[width=86.mm]{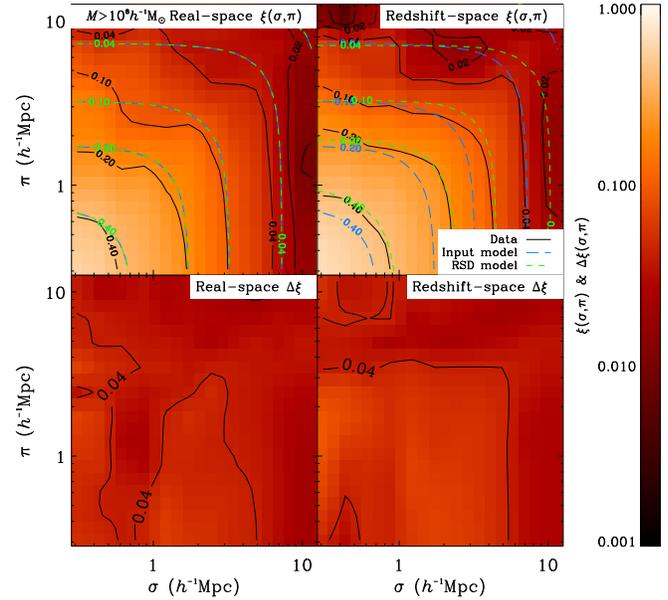} 
\caption{The top two panels show the GIMIC galaxy-Ly$\alpha$ $\xi(\sigma,\pi)$ results (shaded map and solid black contours) based on the \lowmass\ galaxy sample in real-space (top-left panel) and in redshift-space (top-right panel). We show the underlying double power law model derived from the real-space correlation function (i.e. with no RSD modelling) by the long-dashed blue contours (identical in the top two panels). The RSD models that best fit the $\xi(\sigma,\pi)$ results (based on this input model) are shown by the short-dashed green contours. Errors were calculated based on a jack-knife analysis and are shown in the lower panels.}
\label{fig:xisp_lyaLBG_gimic_lm}
\end{figure}

\begin{figure}
\centering
\includegraphics[width=86.mm]{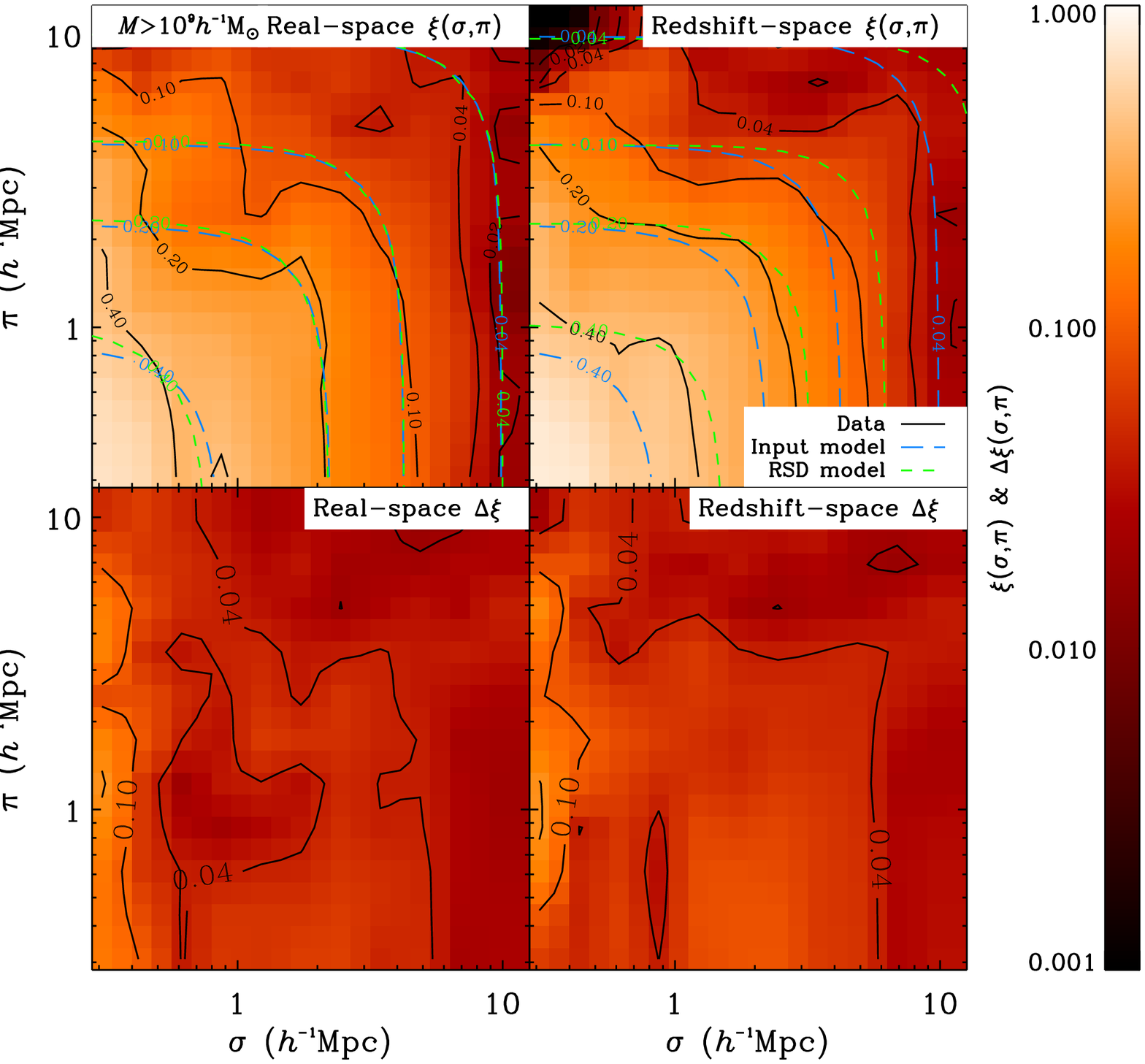}
\caption{As in Fig.~\ref{fig:xisp_lyaLBG_gimic_lm}, but for the \himass\ sample.}
\label{fig:xisp_lyaLBG_gimic_hm}
\end{figure}

We now analyse the properties of the 2-D cross-correlation function, $\xi (\sigma, \pi)$. This is calculated in the same way as $\xi(r)$, whilst again we estimate errors on the results using the jack-knife method. The GIMIC galaxy-Ly$\alpha$ $\xi (\sigma,\pi)$ results are presented in Fig.~\ref{fig:xisp_lyaLBG_gimic_lm} for the \lowmass\ galaxy sample and Fig.~\ref{fig:xisp_lyaLBG_gimic_hm} for the \himass\ sample. In each case the top left panel shows $\xi(\sigma,\pi)$ in real-space and the top right panel in redshift-space, with the lower panels showing the associated error profiles on the same scale. The dashed blue lines show the $r-r$ double power-law fit to the data, which we use as the input model for our RSD model fitting to the $\xi (\sigma, \pi)$ contours, in which we ascertain the best fitting values for the parameters $\beta_{\rm Ly\alpha}$ and $\vdisp$.

\begin{figure}
\centering
\includegraphics[width=80.mm]{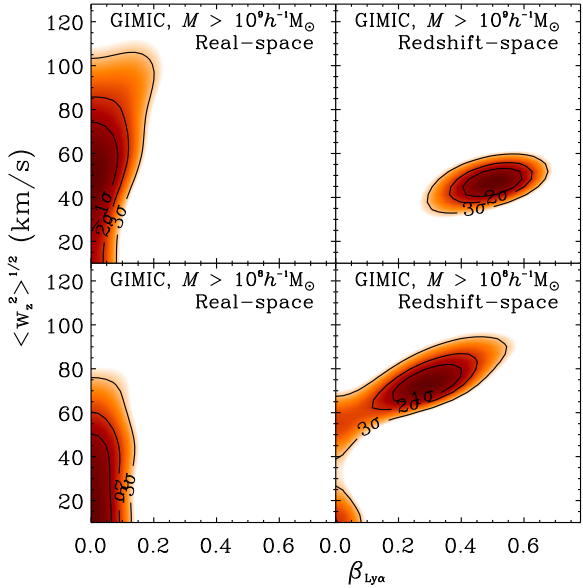}\\
\caption{The $\Delta\chi^2$ contours for the RSD model fitting to the 2D galaxy-Ly$\alpha$ cross-correlation are shown. The top panels show the fitting results for the \himass\ galaxy sample in real-space (left) and redshift-space (right). In real-space, we find best fitting parameters of $\beta_{\rm Ly\alpha}=0.00^{+0.08}_{-0.00}$ and $\vdisp=50\pm25~{\rm km~s}^{-1}$. For the redshift-space $\xi(\sigma,\pi)$ the best fitting parameters are $\beta_{\rm Ly\alpha}=0.51\pm0.12$ and $\vdisp=48\pm7~{\rm km~s}^{-1}$. The lower panels show the results for the \lowmass\ GIMIC sample. The best fit for the real-space sample (left panel) is $\beta_{\rm Ly\alpha}=0.00^{+0.06}_{-0.00}$, $\vdisp=0^{+50}_{-0}~{\rm km~s}^{-1}$. For the redshift-space $\xi(\sigma,\pi)$ we find $\beta_{\rm Ly\alpha}=0.28\pm0.10$, $\vdisp=73\pm9~{\rm km~s}^{-1}$.}
\label{fig:xisp_lyaLBG_gimic_fits}
\end{figure}

We fit to both the real and redshift-space $\xi(\sigma,\pi)$ using the same basic model and limited to a maximum separation of $r=12\hmpc$ (minimising the impact of the limited simulation size on the results). By first fitting to the real-space results, we provide a baseline test of whether the analysis successfully gives $\beta_{\rm Ly\alpha}=0$ and $\vdisp=0~\kps$ for the case of no peculiar velocities. The $\chi^2$ fitting contours for the real-space measurements are shown in the left hand panels of Fig.~\ref{fig:xisp_lyaLBG_gimic_fits} (with the top panel showing the fit to the \himass\ measurement and the lower panel showing the fit to the \lowmass\ measurement). Starting with the \lowmass\ real-space result, we find best fit parameters entirely consistent with the lack of RSD effects on the $\xi(\sigma,\pi)$ measurement, with $\beta_{\rm Ly\alpha}=0^{+0.06}_{-0.00}$ and $\vdisp=0^{+50}_{-0}~\kps$. For the \himass\ measurement, we find $\beta_{\rm Ly\alpha}=0^{+0.08}_{-0.00}$ and $\vdisp=50\pm25~\kps$. These best-fitting models are shown by the green dashed contours in  Fig.~\ref{fig:xisp_lyaLBG_gimic_lm} and Fig.~\ref{fig:xisp_lyaLBG_gimic_hm}. For the \lowmass\ sample, the analysis successfully identifies the lack of any velocity information in the result, however the \himass\ results appears to show a $\sim2\sigma$ signal for a non-zero $\vdisp$, corresponding to some small scale velocity dispersion. The cause of this is evident from the contour plot of $\xi(\sigma,\pi)$ in the top left panel of Fig.~\ref{fig:xisp_lyaLBG_gimic_hm}, where an extension along the $\pi$ axis is clearly visible at small $\sigma$. This extension is at the $\sim2\sigma$ level according to the jack-knife errors and, given that there are no velocity offsets in this realisation, the non-zero result is likely caused by statistical fluctuations at these small $\sigma$ scales. That a zero velocity dispersion is found for the \lowmass\ sample in which we have more galaxies, seems to support this conclusion. However, this is important to factor into the analysis when reapplying the model fitting to the redshift-perturbed simulated galaxies. 

The $\chi^2$ contours for the model fits to the redshift-space $\xi(\sigma,\pi)$ results are shown in the right hand panels of Fig.~\ref{fig:xisp_lyaLBG_gimic_fits} - where the top panels show the results for the \himass\ galaxy sample and the lower panels show the results for the \lowmass\ sample.

For the \lowmass\ galaxies, we find best fitting parameters of $\beta_{\rm Ly\alpha}=0.28\pm0.10$ and $\vdisp=73\pm9~\kps$, whilst for the \himass\ sample, we find $\beta_{\rm Ly\alpha}=0.51\pm0.12$ and $\vdisp=48\pm7~\kps$. The first thing to note is that the measurements for $\beta_{\rm Ly\alpha}$, which should be the same given they both represent the gas motion, are consistent at the $\approx1\sigma$ level between the two samples. Further to this, we can compare to our results from the $\xi(r)$ analysis as a consistency check of the analysis. From $\xi(s)/\xi(r)$, we measured values of $\beta_{\rm Ly\alpha}=0.27\pm0.05$ and $\beta_{\rm Ly\alpha}=0.31\pm0.07$ from the \lowmass\ and \himass\ samples respectively. Collating all the measurements of $\beta_{\rm Ly\alpha}$ thus far then, the $\beta_{\rm Ly\alpha}$ values are all in strong agreement between the $\xi(r)$ \lowmass\ and \himass\ results and the $\xi(\sigma,\pi)$ \lowmass\ result, whilst the $\xi(\sigma,\pi)$ \himass\ result shows some small tension at the $\sim1.5\sigma$ level.

\begin{table}
\caption{Results for the power-law fits to the 2D galaxy-Ly$\alpha$ cross-correlation functions.}
\centering
\begin{tabular}{lcc}
\hline
Sample          & $\beta_{\rm Ly\alpha}$ & $\vdisp$ ($\kps$) \\
\hline
VLRS$+$Keck     & $0.33_{-0.23}^{+0.33}$   & $190\pm90$ \\
GIMIC \lowmass  & $0.28\pm0.10$            & $73\pm9$          \\
GIMIC \himass   & $0.51\pm0.12$            & $48\pm7$          \\
\hline
\end{tabular}
\label{tab:gasgal_2dfits}
\end{table}

Now looking to the velocity dispersion results, we find a significant difference between the \lowmass\ and \himass\ $\xi(\sigma,\pi)$ results. In itself this is not unexpected, given that the galaxy population contributes to this parameter. However, we would expect the \himass\ sample to show a higher velocity dispersion than the \lowmass\ sample, which is not the case. In addition, the \himass\ redshift-space measurement is itself consistent with the \himass\ real-space measurement, suggesting that we are not actually able to measure the velocity dispersion in this case. Given the small number of pairs at small separations in the \himass\ galaxy-Ly$\alpha$ cross-correlation, this is likely due to the stochastic nature of the signal we are measuring at these small scales. As discussed in Section~\ref{comotion}, based on the measured galaxy and gas velocity distributions we would expect $\vdisp \approx 139~\kps$ and $\vdisp \approx 156~\kps$ for the \lowmass\ and \himass\ samples respectively. This is clearly not the case from our measurements. The explanation may be due to the gas motion being very coherent with the galaxies at small scales. Should the gas and galaxies be moving together in such a way, this could reduce the measured velocity dispersion between the two in the cross-correlation analysis - for $\xi(s)$ as well as $\xi(\sigma,\pi)$. The fitting results are summarised in Table~\ref{tab:gasgal_2dfits}.

\subsubsection{Simulation and observation compared}

We next compare the simulated results for galaxy-Ly$\alpha$ $\fltrans$ with the Keck+VLRS data as shown in the right hand panel of Fig.~\ref{fig:xcor_lyalbg_data}. The GIMIC result for the \himass\ sample in redshift space is shown by the shaded green curve, which follows the observational data points well. The GIMIC result falls to lower values of $\fltrans$ at small scales than the observational result, although only at the $\approx1\sigma$ level, a potential sign of the effect of observational velocity errors on the data points.

We compare the simulated results for galaxy-Ly$\alpha$ $\xi(\sigma,\pi)$ now with the 6 Keck quasars $+$ VLRS data as shown in Fig.~\ref{fig:xisplyaLBG}. We have seen that the observed best fit parameters for the Keck $+$ VLRS data are $\beta_{\rm Ly\alpha}=0.33^{+0.33}_{-0.23}$, $\vdisp = 190\pm90~\kps$, with this measurement of the infall parameter being $\approx3\sigma$ lower than the predicted value of $\beta_{\rm Ly\alpha}=1.3$. As we have shown above, the best fit value for $\beta_{\rm Ly\alpha}$ from the simulated galaxy samples covers a range of $\beta_{\rm Ly\alpha}\approx0.3-0.5$. This simulated $\beta_{\rm Ly\alpha}$ value is consistent within the error estimates on our VLRS observations, but not the theoretically motivated $\beta_{\rm Ly\alpha}=1.3$. We also note  that both the minimum value of $\vdisp = 297~\kps$ from LBG velocity error and the $\vdisp = 320~\kps$ value from including the full simulated  $\vdisp = 120~\kps$ are consistent with the data at the $\approx1-2\sigma$ level. We conclude that while the Keck+VLRS $\beta_{\rm Ly\alpha}$ and $\vdisp$ estimates are low compared to initial expectation from theory, they are consistent with the similarly low values of these parameters estimated from $\xi(r)$ and $\xi(\sigma,\pi)$ in the GIMIC simulations.

\section{Auto-correlation analysis of the IGM}
\label{sec:lya_auto}

\subsection{Ly$\alpha$ 1D auto-correlation function}

We now measure the Ly$\alpha$ auto-correlation function in both the observational data and the simulated sightlines, with the aim of again comparing the simulated sightline results to observations and measuring the effect of the velocity field on the clustering. Following \cite{Crighton2011}, for each pixel in a quasar line-of-sight, we calculate:

	\begin{equation}
	\label{delT}
		\delta =\frac{T}{\bar T} -1,
	\end{equation}
	
\noindent where $T$ and $\bar T$ are the measured and the mean normalised flux. We then use this to calculate the auto-correlation function:
	
	\begin{equation}
	\label{xirforlyalya}
		\xi(\Delta s) = \langle \delta(s)\delta(s + \Delta s) \rangle,
	\end{equation}
	
For the observational data, we are only able to do this in the line of sight direction as there are only 3 pairs of quasars that can provide transverse separation measurements and these are all separated by $\gtrsim20\hmpc$. For the simulated sightlines, we sum all pixels with the separations $\Delta s$, both parallel and perpendicular to the line of sight.
	
	\begin{figure}
	\centering
 	\includegraphics[width=80.mm]{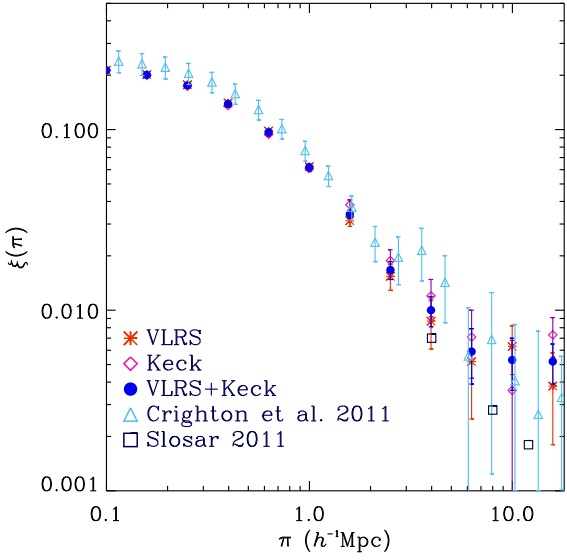}
	\caption{The auto-correlation of Ly$\alpha$ pixels along the line-of-sight. The VLRS, Keck and combined samples are shown by red asterisks, pink diamonds and filled blue circles respectively. The measurement of  \citet{Crighton2011} is also shown (cyan triangles), as is the BOSS result of  \citet[][black squares]{Slosar2011}.}
	\label{fig:xis_lyalya_obs}
	\end{figure}
	
Fig.~\ref{fig:xis_lyalya_obs} shows the auto-correlation of
Ly$\alpha$ pixels along the line-of-sight from the observational data.
Keck, VLRS and combined samples are presented by pink diamonds, red
asterisks and filled blue circles respectively. Error bars were
estimated by using the jack-knife method. We first compare these to the result from
\citealt{Crighton2011} (cyan triangles) who measured the auto-correlation using 7 high
resolution quasars (resolution FWHM $\sim$ 7~km~s$^{-1}$). They all show similar results at small
scales.  We also show the recent BOSS result of \citealt{Slosar2011} (black squares), which probes scales of $\gtrsim3\hmpc$ and is consistent with the VLRS results.

\begin{figure}
	\centering
	\includegraphics[width=80.mm]{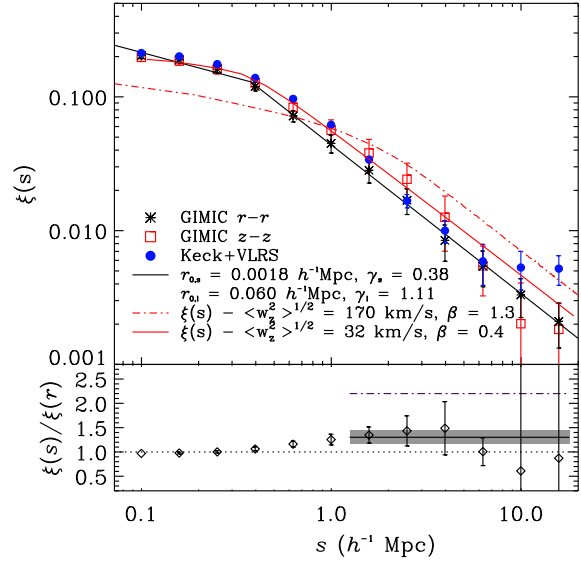}	
	\caption{Top panel: The auto-correlation functions of GIMIC Ly$\alpha$ pixels at $z$ = 3.06 in the  $0\sigma$ simulation. Real-space (black asterisks) and redshift-space (red squares) results are shown. Errors are calculated via the jack-knife method. A double power-law fit to the real-space $\xi(r)$ with $r_{0,s}=0.0018\pm0.0015$, $\gamma_{s} = 0.38\pm0.14$, $r_{0,l}=0.060\pm0.034$, $\gamma_{l}=1.11\pm0.21$ is also shown (black line). The red dot-dashed line is the expected result for the Ly$\alpha$ $\xi(s)$ in redshift-space if we convolve in the velocity dispersion $\vdisp = 170~\kps$ and $\beta_{\rm Ly\alpha}$ = 1.3 to the RSD model. The red solid curve is a model $\xi(s)$ fitted to the GIMIC $z-z$ result and is given by parameter values of $\vdisp = 32~\kps$ and $\beta_{\rm Ly\alpha}=0.40$ (applied to the double power-law model fit to $\xi(r)$). Bottom panel: GIMIC $\xi(s)/\xi(r)$ with jack-knife error bars. The dot-dashed line corresponds to $\xi(s)/\xi(r)=2.2$ as predicted from linear theory with $\beta_{\rm Ly\alpha}= 1.3$, whilst the solid black line and grey surround shows the weighted mean $\xi(s)/\xi(r)$ from the simulation and it's $1\sigma$ bounds.}
	\label{fig:xis_lyalya_gimic}
	\end{figure}

The auto-correlation functions based on the GIMIC simulated Ly$\alpha$ sightlines are presented in Fig.~\ref{fig:xis_lyalya_gimic}. The real and redshift-space Ly$\alpha$ auto-correlation functions are shown by black asterisks and red squares respectively. The VLRS$+$Keck result from Fig.~\ref{fig:xis_lyalya_obs} is replotted (filled blue circles) and is found to be consistent with the GIMIC auto-correlation within the quoted error estimates.

Focussing on the simulation, we see again that the redshift and real-space correlation functions are comparable in amplitude and form. We fit the GIMIC real-space auto-correlation function with a double power law form as performed with the galaxy-Ly$\alpha$ cross-correlation. The resulting fit is given in Fig.~\ref{fig:xis_lyalya_gimic} (solid black curve). At small scales, convolving this double power-law fit with a Gaussian of width $120\times\sqrt{2}=170$ km~s$^{-1}$ representing the simulation gas peculiar velocity (see Fig. \ref{fig:gal-vz}) is seen to overestimate the small-scale turnover in the redshift-space correlation function.

Based on the power-law fit at $r>0.4\hmpc$ (and using the relation $b=\sqrt{\xi_{\rm Ly\alpha}/\xi_{\rm DM}}$), the clustering bias of the Ly$\alpha$ forest is $b\approx0.3$. Assuming $\beta_{\rm Ly\alpha}=\Omega^{0.6}/b$, this bias corresponds to $\beta_{\rm Ly\alpha}\approx3.3$ which implies $\xi(s)/\xi(r)\approx5.4$. But again as noted by \citet{McDonald2003},
$\beta_{\rm Ly\alpha}$ has no simple relation to density bias as for
galaxies. $\beta_{\rm Ly\alpha}$ has to be estimated from simulations and
the simulations of McDonald et al implied a range $\beta_{\rm Ly\alpha} =
1-1.6$. If  we therefore take $\beta_{\rm Ly\alpha} = 1.3$, then this
predicts $\xi(s)/\xi(r) = 2.2$ from Eq.~\ref{kaiser} whereas the
simulated value in Fig.~\ref{fig:xis_lyalya_gimic} is $\xi(s)/\xi(r)
= 1.30\pm0.14$ (solid black line and shaded region in the lower panel of Fig.~\ref{fig:xis_lyalya_gimic}), which corresponds to $\beta_{\rm gal}=0.40\pm0.16$. With $\beta_{\rm Ly\alpha} = 0.40$, the best fit
velocity dispersion is $\vdisp = 32\pm7~\kps$. Models where we
fixed $\beta_{\rm Ly\alpha} = 1.3$ and took $\vdisp = 170~\kps$
as expected from Fig.~\ref{fig:lya_vel} are strongly rejected (red dash-dot curve).
With $\beta_{\rm Ly\alpha} = 1.3$, a best fit value of $\vdisp
= 86^{+11}_{-8}~\kps$ was found although the model was still  rejected in a
chi-square test.

Whatever value of  $\beta_{\rm Ly\alpha}$ is chosen, it appears that the fitted value of the velocity dispersion is much lower than we measured in Fig.~\ref{fig:lya_vel}. However, as shown by \cite{Crighton2011}, the intrinsic width of the Ly$\alpha$ lines convolved with the instrumental response of the spectrograph can induce artificial autocorrelations at scales $\la0.7$~$h^{-1}$Mpc, so this effect may contribute to the poor fit of the peculiar velocity RSD model on small scales.

We note that the RSD model for the Ly-$\alpha$ auto- and cross-correlation assumes spherical symmetry  as we move from real-space to redshift-space and the Ly$\alpha$ auto-correlation function involves summing along and across quasar lines of sight which may not be exactly spherically symmetric. However, we shall see that this explanation cannot apply to the Ly$\alpha$ $\xi(\sigma,\pi)$ which we calculate next and which gives consistent results with the $\xi(s)$ analysis. 

\begin{figure}
\centering
\includegraphics[width=80.mm]{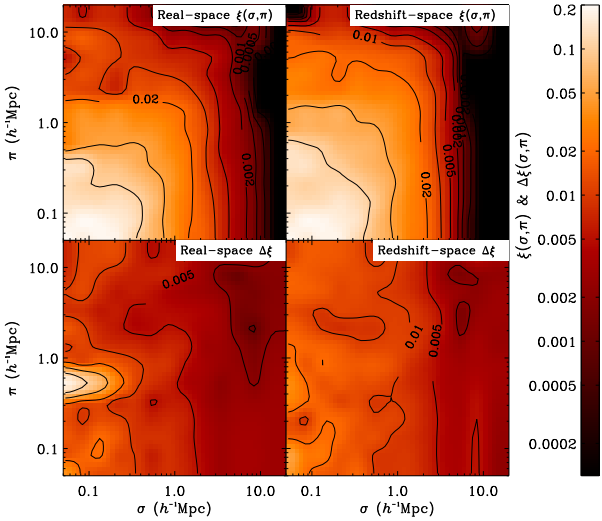}
\caption{The GIMIC Ly$\alpha$ $\xi(\sigma,\pi)$ auto-correlation at $z=3.06$ in real (top-left panel) and redshift-space (top-right panel). The lower panels show the corresponding jackknife error estimates on the $\xi(\sigma,\pi)$ results.}
\label{fig:xisp_lyalya_gimic}
\end{figure}

\subsection{Ly$\alpha$ 2D auto-correlation function}

For each pixel in the Ly$\alpha$ line-of-sight, we next calculate the Ly$\alpha$ $\xi (\sigma, \pi)$ by using,

\begin{equation}
\xi(\sigma,\pi)=\frac{\left<DT (\sigma, \pi)\right>}{N (\sigma, \pi)},
\end{equation}	

\noindent where $\left<DT (\sigma, \pi)\right>$ is the number of Ly$\alpha$ pairs weighted by the normalised transmissivity, $T$, for each separation. N($\sigma, \pi$) is the number of Ly$\alpha$ pixels that contributed to each pair.

The Ly$\alpha$ $\xi(\sigma,\pi)$ results at $z = 3.06$ for the $0\sigma$ simulation are shown in Fig.~\ref{fig:xisp_lyalya_gimic} with the top-left panel showing the result in real-space and the top-right panel showing the result in redshift-space. The associated errors are again shown in the lower panels.

\begin{figure}
\centering
\includegraphics[width=80mm]{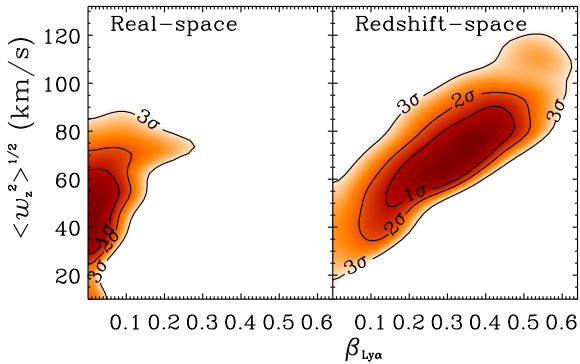}
\caption{Results for the model fits to the GIMIC 2D Ly$\alpha$ auto-correlation functions shown in Fig.~\ref{fig:xisp_lyalya_gimic}. The left panel shows the $\Delta\chi^2$ contours for fit to the real-space $\xi(\sigma,\pi)$ with best fitting parameters of $\beta_{\rm gal}=0.00^{+0.09}_{-0.00}$ and $\vdisp=46\pm17~{\rm km~s}^{-1}$. The right panel shows the same for the redshift-space auto-correlation function. Here the best fit is $\beta_{\rm gal}=0.31\pm0.17$, $\vdisp=69\pm21~{\rm km~s}^{-1}$. The fits are based on the underlying double power-law function shown in Fig.~\ref{fig:xis_lyalya_gimic}.}
\label{fig:betavdisp_lyalya_gimic}
\end{figure}

Again we fit the RSD model to the GIMIC results and find $\vdisp=46\pm17~\kps$ and $\beta_{\rm gal}=0.00^{+0.09}_{-0.00}$ for the real-space result (see left hand panel of Fig.~\ref{fig:betavdisp_lyalya_gimic}). For redshift-space the best fitting parameters are the same with $\beta_{\rm gal}=0.31\pm0.17$ and $\vdisp=69\pm21~\kps$ (see right hand panel of Fig.~\ref{fig:betavdisp_lyalya_gimic}). We again conclude that the effects of infall in the gas in the GIMIC simulation are much less than predicted from the previous work of \citet{McDonald2003} with an upper limit of $\beta_{\rm Ly\alpha}\lesssim0.6$ from $\xi(s)$ and $\beta_{\rm Ly\alpha}\lesssim0.5$ from  Ly$\alpha$ $\xi(\sigma,\pi)$. Given $\beta_{\rm gal}=0.31$, the gas velocity dispersion fit of $\vdisp=69\pm21~\kps$ is close to the sub-$1h^{-1}$Mpc value of the velocity dispersion estimated for simulated galaxies, due to correlated  motions.

As discussed, comparing the VLRS$+$Keck $\xi(s)$ result (filled blue circles) with the GIMIC $\xi(s)$, we found good agreement between the two within the $1\sigma$ errors on the two datasets. However, we do not calculate the Ly$\alpha$ 2-D autocorrelation from the observations as the quasar sample does not have a high enough sky density to probe the on-sky projected profile.

\section{Discussion}

We have combined the power of the VLRS at large spatial scales with the statistical power of the Keck sample at smaller scales. \cite{Crighton2011} included the Keck data in the LBG-Ly$\alpha$ cross-correlation function by simply using an error weighted combination of  the Keck and VLRS correlation functions. Our aim here was to combine the two surveys for 2-D, $\xi(\sigma,\pi)$ correlation function analyses at the deeper level of the Ly$\alpha$ fluxes and LBG positions. We therefore included 940 $2.67\leq z\leq 3.25$ LBGs from the \cite{Steidel2003} Keck samples. We also re-reduced 6 high resolution spectra of the quasars in these fields from the ESO and Keck archives. With $\approx3000$ galaxies the combined VLRS and Keck surveys covering the widest wide range of spatial scales are ideal to study  the dynamical relationship between galaxies and the IGM at $z\approx3$.

We have also incorporated the GIMIC SPH simulation into our analysis in order to aid the interpretation of  the correlation function results. GIMIC was used to create synthetic Ly$\alpha$ spectra and galaxies. We study both galaxy clustering and the relationship between gas and galaxies via the auto- and cross-correlation functions in both 1-D and 2-D.

We have compared the simulated galaxy-galaxy results in real- and redshift-space. The simulated galaxy auto-correlation functions, $\xi(r)$ and $\xi(s)$ (i.e. in real and redshift space), are consistent with being power laws at scales of $r\gtrsim2\hmpc$. At small distances ($r\lesssim1\hmpc$), the LBG-LBG $\xi(s)$ tends to have lower clustering than $\xi(r)$ in real-space, while at larger scales the LBG-LBG $\xi(s)$ results have higher clustering. Qualitatively this is as expected from `finger-of-God' effects at sub 1~$h^{-1}$Mpc scales and dynamical infall at larger scales, characterised by the `Kaiser boost'. Quantitatively, the large scale Kaiser boost for the galaxies is marginally lower than predicted based on the galaxy bias, but only by $\approx1-2\sigma$. At smaller scales the peculiar velocity dispersion measured in the simulation overestimates the difference between real and redshift-space correlation functions. Similar results have been found by \cite{Taruya2010} who found that at high redshift fitting finger-of-god damping terms, as we do here, tended to underestimate the peculiar
velocity dispersion predicted by linear theory. Certainly, a `local'
velocity dispersion measured relative to galaxy pairs  with separations
$<1$~$h^{-1}$Mpc produces improved agreement.

From the simulated galaxy 2D auto-correlation function, $\xi(\sigma,\pi)$, we find values for the galaxy infall parameter, $\beta_{\rm gal}$, consistent within $\approx1-1.5\sigma$ with what would be expected from the measured galaxy bias. The same is seen for the pairwise velocity dispersion. Overall, our RSD model is successful in retrieving the properties of the galaxy velocity field when applied to the clustering measurements from the simulation, whilst conversely the simulation is shown to reproduce a realistic galaxy velocity field well.

Following the galaxy auto-correlation analysis, we performed an analysis of the galaxy-gas cross-correlation. We first analysed the LBG-Ly$\alpha$ $\fltrans$ 1D cross-correlation
function as calculated directly from quasar sightline spectra and LBG positions from the VLRS and Keck surveys. We have re-analysed a subset of 6 fields from the 8 used in the work of A03 and found good agreement, observing the small scale upturn reported in the original work. Our VLRS results on the other hand, agree with those of A05 (and \citealt{Rakic2012}), rather than those of A03 - i.e. a continuous decrease in flux transmissivity around the LBG with no evidence for a spike in transmissivity. \cite{Crighton2011} noted that such a spike could still be present but smoothed away by the errors in the LBG velocities, but this now seems unlikely given the results presented here and those of A05 and \citealt{Rakic2012}.

The inclusion of the gas component along with the galaxies allows us to investigate the effects of gaseous infall on the galaxy-gas distribution. Fitting an RSD model to the observed VLRS$+$Keck LBG-Ly$\alpha$ $\xi(\sigma,\pi)$, we found best fitting parameters of $\beta_{\rm Ly\alpha}=0.33^{+0.33}_{-0.23}$ and $\vdisp=190\pm90~\kps$. The large-scale infall measurement is significantly lower than that predicted by \citet{McDonald2003}, i.e. $\beta_{\rm Ly\alpha}=1.3\pm0.3$, whilst the velocity dispersion measurement is consistent (although lower by $\sim1\sigma$) with the velocity errors on our galaxy redshifts. Interestingly, the second point here leaves little room for any intrinsic velocity dispersion between the gas and galaxies at small scales.  We see similar results when analysing the simulated galaxy-Ly$\alpha$ cross-correlation. Again, we find $\beta_{\rm Ly\alpha}\approx0.3$, whilst the velocity dispersion is measured to be somewhat lower than what we may expect for the gas-galaxy velocity dispersion based on their directly measured individual velocity profiles (i.e. from Fig.~\ref{fig:lya_vel}). Indeed, these small measurements of the galaxy-gas velocity dispersion in both our observations and simulations, may be indicative of highly coherent motion between gas and galaxies at small scales.
	
From the Ly$\alpha$ auto-correlations $\xi(r)$ and $\xi (\sigma, \pi)$,
we see similar results with again small differences between real- and
redshift-space. At small scales, the velocity dispersion needed to fit
the simulated $\xi(s)$ is less than measured directly in the simulation,
although this may be partly explained by the intrinsic width of the
Ly$\alpha$ lines contributing artificial autocorrelation below
separations of $\la0.7$~$h^{-1}$Mpc. At larger scales, the value of
$\xi(s)/\xi(r)$ gives $\beta_{\rm Ly\alpha}=0.4\pm0.16$ rather than the range given by \citet{McDonald2003}, $\beta_{\rm Ly\alpha} \approx 1-1.6$ (but entirely consistent with the results from the GIMIC cross-correlation results). 

At larger scales, one possibility to explain  the low gas infall rate may be due to the presence of feedback in the GIMIC simulations. Galaxy-wide winds powered with initial velocities of 600~km~s$^{-1}$ are invoked in the GIMIC simulations and this is a significant amount since this corresponds to 6 $h^{-1}$Mpc. These winds are modelled by each star particle that forms, imparting a randomly directed 600~km~s$^{-1}$ kick to 4 of its gas particle neighbours. It is possible that this outflow of the gas could
cancel out some of the expected gravitational infall particularly in the
neighbourhood of a galaxy. However, it remains to be seen whether enough
gas particles are outflowing to explain  the lack of infall in the gas
cross- or auto-correlation functions. If the effects of gas outflow
were detectable in the gas dynamics this could be a powerful probe,
since there is no evidence of feedback from any spike in
transmission due to lower neutral gas density close to the galaxy.


Studies by \citet{Rakic2012} and \citet{2013MNRAS.433.3103R} presented the LBG-H{\sc i} cross-correlation at $z\sim2.4$ with observations and simulations respectively. In both cases, the authors report a significant measurement of RSD, showing evidence for both small scale peculiar velocity effects and large scale bulk motion of gas in-falling onto observed and simulated galaxies. \citet{2013MNRAS.433.3103R} find that in terms of the reported large scale `flattening', the observations of \citet{Rakic2012} are consistent with the simulation results for galaxy samples selected with minimum halo masses of $\rm{log}(M_{min}/M_{\odot})=11.6\pm0.2$. This is consistent with the halo masses (measured from galaxy clustering) of \citet{Trainor2012}, but is significantly higher than the halo masses of the galaxy samples used here (and those of \citealt{2013MNRAS.430..425B},A03 and A05). We are unable to probe this larger halo-mass constraint given the size limitations of GIMIC, however we note that our GIMIC 2D cross-correlation results appear qualitatively consistent with the results of \citet{2013MNRAS.433.3103R} at lower minimum halo masses. Additionally, \citet{2013MNRAS.433.3103R} compared their measurements for different feedback prescriptions, finding that including AGN weakened the absorption by H{\sc i} (within $\sim1$ Mpc), whilst increasing the wind mass-loading increased the measured absorption. The authors do not make any quantitive analysis of the effect of increasing the wind mass-loading on the presence of large-scale infall in the cross-correlation analysis. However inspecting their Fig.~4, it is evident that there is indeed some movement in the large scale measurement of the gas distribution when the wind mass-loading is increased (i.e. comparing the `REF' model result to the `WML4' result). This provides some additional motivation for the supposition that SNe driven winds could affect our measurement of $\beta_{\rm Ly\alpha}$. An important test of this will be to apply our RSD modelling to a range of simulation runs incorporating different feedback prescriptions.

\citet{Rakic2012} also investigate the effect of small scale random peculiar velocities on their observed cross correlation, finding evidence for peculiar velocities between gas and galaxies of $\sim240~\kps$. Such a large peculiar velocity is not apparent in the simulation results of \citet{2013MNRAS.433.3103R} and neither is it in our simulation results. Our observations give a consistent measurement of the velocity dispersion with that reported by \citet{Rakic2012}, however this is largely dominated by galaxy redshift errors as we have discussed. 

The \citet{Rakic2012} results have since been further developed by \citet{2014arXiv1403.0942T}, in which increased numbers of the galaxy sample have been observed using the MOSFIRE instrument at the Keck Observatory (improving redshift accuracies). \citet{2014arXiv1403.0942T} report consistent results with \citet{Rakic2012} for the galaxy-H{\sc i} cross-correlation, seeing the same finger-of-god and large-scale infall effects with their improved redshift errors. It is difficult to make a quantitive comparison between our results and those of \citet{Rakic2012,2013MNRAS.433.3103R} and \citet{2014arXiv1403.0942T} however, given the different measurements made. Qualitatively these complementary studies are consistent with the work presented here in identifying the presence of both large-scale infall and small scale peculiar velocity effects in the H{\sc i} gas around $z=2-3$ star-forming galaxies.


A more direct comparison can be made with the results of \citet{Slosar2011}, who measure the $\beta_{\rm Ly\alpha}$ parameter from the auto-correlation of the Ly$\alpha$ forest in BOSS quasar spectra. They find a range of $0.44<\beta_{\rm Ly\alpha}<1.20$, at central redshift $z = 2.25$. This large range is however consistent at the $1\sigma$ level with all of the other results considered, i.e. the VLRS$+$Keck observations, the GIMIC simulation results and the theoretical prediction from \citet{McDonald2003}. Interestingly though, assuming $\beta_{\rm Ly\alpha}$ behaves as $\beta_{\rm gal}$, then it should decrease with increasing redshift and we would expect the $z=3$ result to be marginally lower than $\beta_{\rm Ly\alpha}$ at $z=2.25$, which is what we find in our study.

\section{Conclusions}
     	
We have analysed the interaction between galaxies and the IGM using a large sample of $z\sim3$ LBGs, in combination with spectroscopic observations of background quasars. In addition to the observational data, we employ the SPH GIMIC simulation to analyse the clustering of gas and galaxies.
	
1. We analyse the auto-correlation of simulated galaxies in the GIMIC simulation using two samples: \lowmass\ and \himass. The \himass\ sample was chosen to match the clustering amplitude of observed LBGs, while the \lowmass\ sample provides a comparison set with higher numbers and hence better statistics. In the simulated data the difference between the real and redshift-space correlation functions
is too small to be self-consistently explained by the measured  peculiar
velocity distribution. We suggest that this is the consequence of a scale dependence in the measurement of the peculiar motions and that the peculiar motions taken within $\lesssim1~h^{-1}$Mpc give a more consistent result.

2. We have checked for the existence of the transmission spike near
star-forming galaxies in the data and  GIMIC simulations which could be
indicative of the effects of star-formation feedback on the IGM. For the
data, we combined the full VLRS and Keck LBG-Ly$\alpha$ datasets to
study both $\xi(r)$ and $\xi (\sigma, \pi)$ and the LBG-Ly$\alpha$
correlation functions. We find no evidence for a transmission spike at small scales and instead find that the gas transmissivity monotonically drops towards the galaxy, consistent with the density of neutral gas rising towards the galaxy position. Although the simulation transmission rises when  LBG velocity errors are taken into account, the simulated and observational results remain in good
statistical agreement.

3. The redshift-space galaxy-Ly$\alpha$ cross-correlation function in the
simulation is close to the real-space correlation function and to some
extent this is  predicted from linear theory applied to the Ly$\alpha$
forest flux which has  a non-linear relation with optical depth and thus
implies lower rates of dynamical infall of gas into galaxies than would
otherwise apply. We have also considered whether galaxy-wide outflows may
be  cancelling out the infall effect.
	
4. The observed Ly$\alpha$ autocorrelation function is also consistent with the simulation. At small scales the difference between real and redshift-space correlation functions in the simulation is again less than  predicted given the peculiar velocity distribution. At larger scales, we measure the effects of dynamical infall and find them to be less than predicted based on the simulations of \citet{McDonald2003}. This may be the residual effect from gas outflows cancelling out the effects of dynamical infall.

5. In the simulations, both gas and galaxies show evidence of a strong bulk motion. 
This bulk motion is undetectable by observable correlation functions but 
may have a connection with the local coherence needed to explain why 
distribution of peculiar velocities overestimates the finger-of-God effect.

\section*{Acknowledgements}

We would like to thank Cai Yanchuan for the useful discussions. PT acknowledges financial support from the Royal Thai Government. TS and RMB would like to acknowledge the funding of their work through the UK STFC research grants. Simulations were performed on the ICC Cosmology Machine, which is part of the DiRAC Facility jointly funded by STFC, the Large Facilities Capital Fund of BIS, Durham University and by the Inter-university Attraction Poles Programme initiated by the Belgian Science Policy Office ([AP P7/08 CHARM]).

\bibliographystyle{mnras_mod}
\bibliography{ref}

\end{document}

%% file: format.tex
\def \hmpc{~h^{-1}\,{\rm Mpc}}

\def \xir{\xi(r)}

\def \gsim { \lower .75ex \hbox{$\sim$} \llap{\raise .27ex \hbox{$>$}} }
\def \lsim { \lower .75ex \hbox{$\sim$} \llap{\raise .27ex \hbox{$<$}} }